\documentclass[twocolumn]{revtex4}
\usepackage{amsmath,amssymb,graphicx,bbm,comment,subfigure}
\newcommand{\mathi}{\mathrm{i}}
\newcommand{\tmop}[1]{\ensuremath{\operatorname{#1}}}
\newcommand{\um}{-}

\begin{document}

\title{Adjacent Spin Operator Dynamical Structure Factor of the $S=1/2$ Heisenberg Chain}

\author{ Antoine Klauser${}^{1,2}$, Jorn Mossel${}^2$, Jean-S\'ebastien Caux${}^2$}

\affiliation{${}^1$Instituut-Lorentz, Universiteit Leiden, P. O. Box 9506, 2300 RA Leiden, The Netherlands}
\affiliation{${}^2$Institute for Theoretical Physics, Universiteit van  
Amsterdam, P. O. Box 94485, 1090 GL Amsterdam, The Netherlands}
\date{\today}

\begin{abstract}
Considering the adjacent spin operators $S^z_jS^z_{j+1}$ and $S^-_jS^-_{j+1}$ in the $S=1/2$ Heisenberg chain, we give a determinant representation of their form factors. The dynamical structure factors of the respective operators are computed over the whole Brillouin zone in several magnetic fields and the resulting signal is analyzed in terms of excitation types. Among other results, we find that the $S^z_jS^z_{j+1}$ dynamical structure factor carries a large weight of the $4$-spinon excitations which are distinguishable from the $2$-spinon signal because they are located outside the $2$-spinon spectrum.
\end{abstract}

\maketitle
\tableofcontents
\section{Introduction}
One-dimensional (1D) systems are particularly prone to quantum criticality as compared to higher-dimensional ones. They offer a prolific playground for low-temperature phenomenology
\cite{giamarchibook,sutherlandbook}. Realizations of 1D critical quantum liquids are numerous and can take the form of e.g. stripes in cuprate high-temperature superconductors, trapped ultracold atomic gases and carbon nanotubes.
One of the simplest strongly correlated models, which remains of great interest due to its underlying richness, is the Heisenberg $S=1/2$ antiferromagnetic chain~\cite{1928_heisenberg_zp_49} 
with Hamiltonian given by
\begin{equation}
H=J\sum_{j=1}^N\frac{1}{2}\left(S^-_jS^+_{j+1} + S^+_jS^-_{j+1}\right)+\left(S^z_jS^z_{j+1} -\frac{1}{4}\right) -hS^z_j
\label{eq:Hamiltonian}
\end{equation}
where $N$ is the number of lattice sites, periodic boundary conditions are understood. Several crystals are known to realize 1D spin chain e.g. KCuF$_3$, SrCuO$_3$, CuPzN and have been probed by neutron scattering \cite{1991_nagler_prb,2005_lake_natmat,2003_stone_prl,2004_zaliznyak_prl,2009_walters_natphys_5,2008_nilson_chemm,2005_enderle_epl}. For a positive coupling constant $J>0$ and zero external magnetic field $h=0$ the ground state is antiferromagnetic without long-range order. Its basic excitations are spinons, fractionalized spin excitations that emerge in the critical state~\cite{1981_faddeev_pla_85}. The Heisenberg model is integrable with an exact solution given
by Bethe Ansatz \cite{1931_bethe_zp_71,1958_orbach_pr}. The properties of its ground state have been extensively studied in \cite{1966_yang_pr1,1966_yang_pr2,1966_yang_pr3}. Moreover, in the case where a spin-exchange anisotropy is introduced, one can obtain a complete phase diagram in terms of the magnetic field and the anisotropy parameter \cite{1982_Kurmann_Phy}. The isotropic Heisenberg chain belongs to the universality
class of Luttinger liquids and the asymptotics of its correlation functions can be calculated by effective field theory methods.
 Whereas the computation of correlation functions was for a long time a limitation of the Bethe Ansatz method,
 significant progresses have been made in the last decade \cite{jimbobook,korepinbook} allowing the calculation of single spin operator dynamical structure factors (DSF) in the $S=1/2$ spin chain \cite{1997_karbach_prb_55,1999_kitanine_npb_554,2000_kitanine_npb_567,2003_caux_prb,2005_caux_prl_95,2005_caux_jstat_p09003,2007_pereira_jstat,2008_caux_jstat_p08006,2011_caux_prl}.

Going one step further from what has been studied so far, we consider the calculation of what we shall call the adjacent spin operator DSF
\begin{eqnarray}
&&\hspace{-1cm}S^{ab \bar{a}\bar{b}}(q,\omega) =  \frac{1}{N} \sum_{j,j'=1}^N\int_{-\infty}^{\infty}dt e^{-\mathi q (j-j')+\mathi \omega t} \nonumber\\
&&\hspace{2cm}\langle S^a_j(t)S^b_{j+1} (t)S^{\bar{a}}_{j'}(0)S^{\bar{b}}_{j'+1}(0) \rangle \; .
\label{eq:adjacent_DSF}
\end{eqnarray}
where $a,b=z,\um,+$. Resuming the study of the spin-exchange DSF in \cite{2011_klauser_prl}, our interest lies in the computation of specific components of the latter correlation function. In addition to the resonant inelastic x-ray scattering response function studied in this previous paper, the adjacent spin operator form factors (FF) are related to the study of transport and spin currents along the spin chain \cite{2003_zotos_arx,1998_narozhny_prb,2011_sirker_prb,2011_steinigeweg_prl,2011_hoogdalem_prb} and to the fidelity in the Heisenberg spin chain \cite{2006_zanardi_pre,2010_albuquerque_prb}.

This paper presents a non-perturbative calculation of the adjacent spin operator DSF for $(a,b)=(z,z),(\um,\um)$ (in \eqref{eq:adjacent_DSF}) in the antiferromagnetic $S=1/2$ Heisenberg chain.

Low-energy effective field theories have already been used to predict the asymptotics of the adjacent spin operator correlation function in a few cases \cite{1989_giamarchi_prb_39,2004_hikihara_prb}.

The contents of this publication are the following. In section \ref{sec:Setup}, we present the exact eigenstates of the system provided by the Bethe Ansatz. Using a determinant representation, the FFs of the operators $S^z_jS^z_{j+1}$ and $S^-_jS^-_{j+1}$ are formulated in terms of eigenstates rapidities in section \ref{sec:FFs}. We show in section \ref{sec:DSF_S4z} and \ref{sec:DSF_Smm} how the $S^{--++}(q,\omega)$ and $S^{4z}(q,\omega)$ DSF are constructed using the FF expressions. In section \ref{sec:catalog}, a catalog of the excitations is given by describing the quasiparticles which occur in presence or absence of magnetic field. In section \ref{sec:DSF_eval}, we evaluate numerically the expression for the two DSFs over the entire Brillouin zone. The contribution of each family of excitations is shortly discussed. The results are summarized in the conclusion in section \ref{sec:conclusion}.

\section{Setup}
\label{sec:Setup}
\subsection{Bethe Ansatz}
Thanks to the integrability of the Heisenberg spin chain, the exact eigenfunctions of (\ref{eq:Hamiltonian}) can be obtained via the  Bethe Ansatz \cite{1931_bethe_zp_71}.
Because the Hamiltonian conserves the total magnetization of the chain, the Hilbert space separates into subspaces of fixed magnetization $M=N/2-\langle \sum^N_{j=1} S^z_j \rangle$. 
Defining a reference state with all spins pointing upwards $\lvert 0\rangle=\bigotimes^N_{j=1}\lvert \uparrow \rangle $,
 the Bethe Ansatz yields, for eigenstates  belonging to the $M$-subspace:
\begin{eqnarray}
\lvert\{k\}_M\rangle &=&\sum_{P\in \pi_M}\left(-1\right)^{[P]}\exp\left( \frac{\mathi}{2}\sum_{1\leq a < b \leq M} \varphi\left(  k_{P_a}, k_{P_b}\right) \right)\nonumber\\
&\cdot&\sum_{j_1,\ldots,j_M}\exp\left( \mathi \sum_{a=1}^M k_{P_a}j_a  \right) \lvert j_1,\ldots,j_M \rangle \; .
\label{eq:coord_BA}
\end{eqnarray}
$\{ k \}$ is a set of $M$ quasi-momenta determining the eigenstate and the scattering phase
 $\varphi\left(  k_a, k_b\right)$ is defined by $2\cot\frac{\varphi\left(  k_a, k_b\right)}{2}=\cot\frac{k_a}{2}-\cot\frac{k_b}{2}$. We introduced the state of $M$ localized down-flipped spins as 
\begin{equation}
 \lvert j_1,\ldots,j_M \rangle=S^-_{j_1} \ldots S^-_{j_M}\lvert 0\rangle\; .
\end{equation}
It is convenient to express  the quasi-momenta  in terms of rapidities $\lambda_i$
\begin{equation}
 \exp(\mathi k_i) = \frac{\lambda_i + \mathi/2 }{\lambda_i - \mathi/2 }\; .
\end{equation}
Imposing periodic boundary conditions leads to the Bethe equations for the rapidities 
\begin{equation}
 \arctan \left(2 \lambda_i\right) =\frac{\pi}{N}I_i + \frac{1}{N} \sum_{k=1}^{M}\arctan (\lambda_i-\lambda_k),\quad  \forall i \; .
\label{eq:Bethe}
\end{equation} 
Taking $N$ to be even, the quantum numbers ${I_1,\ldots, I_M}$ are a set of distinct integers for odd $M$ and half-integers for even $M$.
Each of these sets of quantum numbers specifies a set of rapidities and vice versa. It can be shown that the set of quantum numbers for the ground state is $\{ I^0_i=i -\frac{M+1}{2} \},\> i=1,\ldots,M$. Once the rapidities of an eigenstate are obtained, the energy and momentum of the state follow straightforwardly
\begin{eqnarray}
E& =&-J \sum_{i=1}^{M}\frac{1/2}{1/4+\lambda_i^2}-h(\frac{N}{2}-M) \nonumber\\
 P& =&\pi M -\frac{2 \pi}{N}\sum_{i=1}^{M} I_i \>\> (\text{mod}  \> 2\pi).
\label{eq:E&P}
\end{eqnarray}
The Bethe equations represented in (\ref{eq:Bethe}) provide a systematic way of constructing eigenstates and the solutions allow one to obtain dynamical correlation functions and thermodynamic quantities with high precision.

\subsection{String solutions}

The solutions of \eqref{eq:Bethe} are not restricted to real rapidities only. As Bethe mentioned in his seminal work \cite{1931_bethe_zp_71},
 self-conjugate complex solutions, which can be interpreted as bound states of down spins, also satisfy the Bethe equations.
Bethe \cite{1931_bethe_zp_71} made the conjecture, completed later by Takahashi {\it et al.} \cite{1971_takahashi_ptp}, that such complex rapidities form string structures which are 
$\lambda_{\alpha}^{j,a} = \lambda_{\alpha}^{j} + \frac{i}{2} (n_j + 1 - 2 a)+\mathi\delta^{j,a}_\alpha$
 with $a=1,\ldots,n_j$, $n_j$ being the string length and the string deviation $\delta^{n_j,a}_\alpha$. 
 Here $\alpha$ is an index that runs from $1$ to $M_j$ where $M_j$ is the number of $n_j$ strings (strings of length $n_j$) and $j=1,\ldots,N_s$ where $N_s$ is the total number of different possible lengths. The string hypothesis is unfortunately not always correct,
 although it can be shown that generally the string deviations vanish exponentially with system size. The correct number of complex solutions is moreover smaller than the one constructed from the string hypothesis and other kinds of complex solutions appear.
The proportion is of order $1/\sqrt{N}$ \cite{2003_fujita_jpa}.
 More extended discussions about string structures can be found in \cite{2007_hagemans_jpa}.
The deviated string solutions turn out to be unimportant for the correlation functions presented in this paper.

In the presence of string solutions with vanishing deviations the Bethe equations \eqref{eq:Bethe} become undetermined. The remedy is to rewrite the Bethe equations only in terms of the real centers
 $\lambda_{\alpha}^{n_j}$ of the strings to obtain the Bethe-Takahashi equations \cite{1971_takahashi_ptp}:
\begin{equation}
N \theta_{n_j}(\lambda^{j}_\alpha)-\sum_{k=1}^{N_s}\sum_\beta^{M_k=1} \Theta_{n_j,n_k}(\lambda^j_\alpha-\lambda^k_\beta) = 2\pi I^j_\alpha
\label{eq:Bethe-Takahashi_eq}
\end{equation}
where
\begin{eqnarray}
 \theta_{n_j}( \lambda ) & = & 2 \text{atan} \left(2 \lambda / n_j \right) \nonumber\\
\Theta_{n_j,n_k}(\lambda) & =& (1-\delta_{j,k})\theta_{\lvert n_j - n_k \rvert + 2} ( \lambda ) + 2\theta_{ n_j - n_k - 2} ( \lambda ) + \nonumber\\
& + & \ldots + 2\theta_{ n_j + n_k - 2} ( \lambda ) + \theta_{ n_j + n_k} ( \lambda )
\end{eqnarray}
and $ { I^{j}_\alpha } $ are the set of quantum numbers corresponding to strings of length $n_j$. 

\subsection{Spin of the eigenstates and infinite rapidity }
\label{sec:spin_setup}
In the isotropic spin chain the total spin operator $\big(\sum_j \mathbf{S_j}\big)^2$ commutes with the Hamiltonian and in consequence its eigenvalues  $S^{tot}$ are conserved. Moreover, the eigenstates created with the Bethe Ansatz are highest-weight with respect to the global su(2) algebra i.e. the states constructed with $M$ rapidities have the spin eigenvalues $S^{tot}_z=S^{tot}=\frac{N}{2}-M$ \cite{gaudinbook}. In order to access eigenstates with $S^{tot}_z<S^{tot}$, one must act on a Bethe state with the global spin lowering operator $S^-_{q=0}=\frac{1}{\sqrt{N}}\sum_{j=1}^N S^-_{j}$.
These descendant states are actually also Bethe solutions but they include infinite rapidities. Indeed the Bethe equations allow rapidities to go to infinity and from eq.~\ref{eq:coord_BA}, one notices that a state with $M$ rapidities of which $M'$ tend to infinity can be written $\lvert \{\lambda_1,\ldots,\lambda_{M-M'},\underbrace{\infty,\ldots,\infty}_{M'}\}\rangle = \left(S^-_{q=0}\right)^{M'} \left \lvert \{\lambda_1,\ldots,\lambda_{M-M'} \} \right\rangle$. This eigenstate is no longer highest-weight but has $S^{tot}=\frac{N}{2}-M+M'$ and $S_z^{tot}=\frac{N}{2}-M$. 

The norm of a state which contains one or two infinite rapidities can easily be calculated by commutation of spin operators and are
\begin{eqnarray}
 N (\{ \lambda,\infty\}_M)&=&\frac{N - 2 M + 2}{N} N (\{\lambda\}_{M-1})\; ,\\
 N (\{ \lambda,\infty,\infty\}_M)&=&\frac{2 \left( N - 2M +4 \right) \left( N - 2 M +3 \right)}{N^2}\nonumber\\
&&\cdot\; N 
(\{\lambda\}_{M-2})\; .
\end{eqnarray}

\section{Form factor determinant representations}
\label{sec:FFs}
The natural language for expressing normalized eigenstates and form factors (FF) is the algebraic Bethe Ansatz \cite{korepinbook,1999_kitanine_npb_554,2000_kitanine_npb_567}.
  In the algebraic Bethe Ansatz formalism, one introduces the monodromy matrix $T(\lambda)$ that
acts on the space ${\mathbb C}^2\otimes {\mathcal H}_N$ where ${\mathcal H}_N$ is the Hilbert space of the $N$-sites chain.
The operator is represented in the auxiliary space   ${\mathbb C}^2 $ as
\begin{equation}
T(\lambda)= \begin{pmatrix}A(\lambda) & B(\lambda)\\C(\lambda) & D(\lambda)\end{pmatrix}
\end{equation}
where $\lambda$ is called the spectral parameter. The monodromy matrix is constructed as a product of $L-$operators
\begin{equation}
T(\lambda) = L_N(\lambda;\xi_N) \ldots L_1(\lambda;\xi_1)
\end{equation}
with $L_j(\lambda;\xi_j) = R_{0j}(\lambda-\xi_j)$, where $0$ refers to the auxiliary space and $j$ to the $j-$th site of the spin chain. The $R-$matrix $R_{ij}$ must be a solution of the Yang-Baxter equation. The $\xi_j$ are inhomogeneity parameters and the homogeneous spin chain \eqref{eq:Hamiltonian} corresponds to setting all $\xi_j \rightarrow i/2$. However, it is more convenient to take this limit at the end of the computation, as we will do in this work. One can now define the transfer matrix as $\mathcal{T}(\lambda)  = A(\lambda)+D(\lambda)$, from which  the Hamiltonian can be recovered
\begin{equation}
H = \frac{i}{2} \left. \frac{d}{d\lambda} \ln \mathcal{T}(\lambda) \right|_{\lambda = i/2}\; .
\end{equation}
Hence the transfer matrix and the Hamiltonian have common eigenfunctions and can be constructed as
\begin{equation}
B(\lambda_1) \ldots B(\lambda_M) |0\rangle
\end{equation}
and the dual state as
\begin{equation}
\langle 0 | C(\lambda_1) \ldots C(\lambda_M)
\end{equation}
where the rapidities $\lambda_j$ have to satisfy the Bethe equations \eqref{eq:Bethe}. The norm of eigenstates can now be expressed as a determinant \cite{1981_gaudin_prd_23,gaudinbook,1989_slavnov_tmp_79}.
 Another very import result is a determinant expression for the scalar product \cite{1989_slavnov_tmp_79}
\begin{equation}
\langle 0| C(\mu_1) \ldots C(\mu_M) B(\lambda_1) \ldots B(\lambda_M) | 0 \rangle
\label{eq:scalar_prod}
\end{equation}
where the $\mu_j$ can be arbitrary and the $\lambda_j$ have to satisfy the Bethe equations or vice versa. A last essential ingredient is the inverse mapping that expresses the local spin operators $\sigma^z_j, \sigma^+_j, \sigma_j^-$ in terms of the non-local operators $A(\lambda),B(\lambda),C(\lambda),D(\lambda)$ \cite{1999_kitanine_npb_554}

\begin{eqnarray}\label{eq:ABA_sigma_j}
&& \hspace{-0cm}\sigma^z_j = - 2 \prod^{j - 1}_{i = 1} \left[ A + D \right] (\xi_i)
D (\xi_j) \prod^N_{k = j + 1} \left[ A + D \right] (\xi_k) + \mathbbm{1},\;  \nonumber\\
&&\sigma^-_j =  \prod^{j - 1}_{i = 1}  \left[ A + D \right] (\xi_i)
B (\xi_j) \prod^N_{k = j + 1} \left[ A + D \right] (\xi_k),\; \nonumber\\
&&\sigma^+_j =  \prod^{j - 1}_{i = 1}  \left[ A + D \right] (\xi_i)
C (\xi_j) \prod^N_{k = j + 1} \left[ A + D \right] (\xi_k)\; .
\end{eqnarray}
 Using these expressions and the scalar product \eqref{eq:scalar_prod}, we expressed in terms of determinants the adjacent spin operator form factors which are defined as
\begin{equation}
\langle GS\rvert S^a_{j} S^b_{j+1} \lvert \alpha \rangle  
\end{equation}
with $\lvert GS \rangle $, $\lvert \alpha \rangle$ respectively the groundstate and an eigenstate of the system. The form factors involving a single spin operator were first derived in \cite{1999_kitanine_npb_554}, the general case was treated in \cite{2000_kitanine_npb_567}. These expressions are not always suitable for a numerical evaluation, therefore we re-derive below the form factors of interest for this paper.

\begin{widetext}
 
\subsection{$S^z_jS^z_{j+1}$}
With  $\langle \{ \lambda \}|$ and $|\{\mu\}\rangle$ two eigenstates of the system satisfying the Bethe equations,
the form factor of the operator $S^z_jS^z_{j+1}=\frac{1}{4}\sigma_j^z  \sigma_{j+1}^z $ reads
 $\frac{1}{4}\langle \{\lambda\} | \sigma_j^z  \sigma_{j+1}^z  |\{\mu \}\rangle$.
Using \eqref{eq:ABA_sigma_j} and the properties of the transfer matrix, the product of two adjacent $\sigma_j^z$ operators is

\begin{eqnarray}
  \text{$\sigma^z_j$}  \text{$\sigma^z_{j + 1}$} & = & 4 \prod^{j - 1}_{i = 1} \left[ A + D \right] (\xi_i) D (\xi_j) D (\xi_{j +
1}) \prod^N_{i = j + 2} \left[ A + D \right] (\xi_i)\nonumber\\
&& + \text{$\sigma^z_j$} +
\text{$\sigma^z_{j + 1}$} - \mathbbm{1} \; .
\end{eqnarray}
The form factor for a single $\sigma^z_j$ operator has been considered in \cite{1999_kitanine_npb_554,2005_caux_jstat_p09003} and the remaining non-trivial part of $\sigma^z_j \sigma^z_{j+1}$ form factor is the term: $\prod^{j - 1}_{i = 1} \left[ A + D \right] (\xi_i) D (\xi_j) D (\xi_{j +
1}) \prod^N_{k = j + 2} \left[ A + D \right] (\xi_k)$.
Using the expression for the action of the $D(\lambda)$ operator on a general state \cite{korepinbook} twice, we can write the form factor as a double summation of Slavnov determinants

\begin{eqnarray}
\left\langle \{\lambda\} | D (\xi_j) D (\xi_{j + 1}) | \{\mu\} \right \rangle &=&
  \sum^M_{n = 1} d (\mu_n) \prod_{i \neq n} \frac{\phi (\mu_n - \mu_i +\eta)}{\phi (\xi_j - \mu_i)} \sum^M_{m = 1, m \neq n} d (\mu_m) \prod_{i\neq m} \frac{\phi (\mu_m - \mu_i + \eta)}{\phi (\xi_{j + 1} - \mu_i)}\nonumber\\
  &\cdot& \frac{\phi^2 (\eta)}{\phi (\mu_m - \mu_n + \eta)} \frac{1}{b^{} (\mu_m,\xi_{j + 1})} \frac{1}{\prod_{l > k} \phi (\lambda_k - \lambda_l) \prod_{l < k} \phi (\mu_k - \mu_l)}\nonumber\\
  &\cdot& \frac{\det H (\{\lambda_i \}, \{\mu_{i \neq m, n}, \xi_j, \xi_{j + 1}\})}{\phi (\xi_j - \xi_{j + 1})}
\end{eqnarray}
with
\begin{eqnarray}
H_{a b} (\{\lambda\}, \{\mu\}) & = &\frac{\phi (\eta)}{\phi (\lambda_a \um \mu_b)} \left( \prod_{k\neq a} \phi (\lambda_k \um \mu_b + \eta) - d (\mu_b) \prod_{k \neq a} \phi(\lambda_k \um \mu_b - \eta) \right)\nonumber\\
d(\lambda) &=& \prod_{i=1}^N b(\lambda,\xi_i),\quad b(\mu,\lambda) =\frac{\phi(\mu-\lambda)}{\phi(\mu- \lambda+\eta )} \; .
\label{eq:H_matrix}
\end{eqnarray}
The value of $\eta$ and the definition of $\phi(\lambda)$ depend on the anisotropy parameter of the spin chain \cite{1999_kitanine_npb_554}. We keep their general expressions for the rest of this section in order to cover any anisotropy. However, in the isotropic case they take the values
\begin{equation}
\phi(\lambda) = \lambda,\quad \eta=\mathi \; .
\end{equation}
\subsubsection{Homogeneous limit}
The homogeneous spin chain corresponds to the case where $\xi_i \rightarrow \eta/2 \quad \forall i$. This limit should be taken with care in order to obtain a finite expression for the ratio  $\det H (\{\lambda \}, \{\mu_{i \neq m, n}, \xi_j, \xi_{j + 1}\}) /\phi(\xi_{j+1}-\xi_j) $. 
We use l'H\^opital's rule to have a well-defined homogeneous limit and the matrix elements become
\begin{eqnarray}
\langle \{\lambda\}| D (\eta/2) D (\eta/2) |\{\mu\} \rangle &=&
\frac{\prod_i \phi^2 (\lambda_i + \eta / 2) \phi^{- 2} (\mu_i - \eta /
2)}{\prod_{j > k} \phi (\lambda_k - \lambda_j) \prod_{j < k} \phi (\mu_k -
\mu_j)}  \sum_{n = 1}^M A^n \sum_{m = 1}^M \det F^{n m}\; .
\end{eqnarray}
Here we defined
\begin{eqnarray}
  A^n & = & d (\mu_n) \phi (\mu_n - \eta / 2) \prod_{i_{} = 1}^M \phi (\mu_i -\mu_n - \eta)\nonumber\\
  B_{a b}^n & = & (1 - \delta_{b, n}) d (\mu_b) \phi (\mu_b + \eta / 2) \prod_{i\neq n} \phi (\mu_i - \mu_b- \eta)\frac{\phi (\eta)}{\phi (\lambda_a \um \eta / 2) \phi (\lambda_a +
  \eta / 2)}\nonumber\\
C_a & = & \frac{\phi (\eta) \phi(2 \lambda_a)}{\phi^2 (\lambda_a - \eta / 2) \phi^2 (\lambda_a + \eta
  / 2)}\nonumber\\
  F_{a b}^{n m} & = & 
\begin{cases}	
    G^n_{a b}  \;, & b \neq m\\
    B^{n}_{ab}  \;, & b = m
  \end{cases}\nonumber\\
  G^n_{a b}  & = & 
  \begin{cases}
    \frac{\phi (\eta)}{\phi (\lambda_a \um \mu_b)} \left( \prod_{k
    \neq a} \phi (\lambda_k \um \mu_b + \eta) - d (\mu_b) \prod_{k \neq a}
    \phi (\lambda_k \um \mu_b - \eta) \right)\; ,& b \neq n\\
    C_a  \;, &b = n
  \end{cases} \; .
\label{eq:homo_limit}
\end{eqnarray}
Using Laplace's determinant formula (see chapter 9 appendix in \cite{korepinbook}) and the fact that $B^n$ is a rank one matrix, we write the summation over determinants as a single determinant:
\begin{eqnarray}
  \sum_{m = 1}^M \det F^{n m} & = & \det (G^n +B^n) - \det G^n
\end{eqnarray}
and the complete form factor is then
\begin{eqnarray}
\left\langle \{\lambda\} | \sigma^z_j \sigma^z_{j +
1} | \{\mu\} \right\rangle & = & \left\langle \{\lambda\} | \sigma^z_j | \{\mu\} \right \rangle + \left\langle \{\lambda\} | \sigma^z_{j + 1} | \{\mu\} \right\rangle -
  \left\langle \{\lambda\} | \{\mu\} \right\rangle \nonumber\\
 & + & 4 \frac{\varphi_{j - 1} (\{\lambda\})}{\varphi_{j - 1} (\{\mu\})}
  \frac{\prod_i \phi^2 (\lambda_i + \eta / 2) \phi^{- 2} (\mu_i + \eta /
  2)}{\prod_{j > k} \phi (\lambda_k - \lambda_j) \prod_{j < k} \phi (\mu_k -
  \mu_j)}  \sum_{n = 1}^M A_n (\det (G^n + B^n) - \det G^n)\; .
\label{eq:SzSz_FF}
\end{eqnarray}
Here we introduced
\begin{equation}
\varphi_j(\{\lambda\}) = \prod_{l=1}^M \left( \frac{\phi(\lambda_l - \eta/2)}{\phi(\lambda_l + \eta/2)} \right)^j
\end{equation}
and if $\{ \lambda \}$ is an eigenstate, then we can write $\varphi_j(\{\lambda\})=e^{- \mathi j P_{\{ \lambda \} } }$ with $P_{\{ \lambda \}}$ the total momentum of the state.
\subsubsection{Fourier transform}
For the further computation of the dynamical structure factor, the Fourier transform of the
form factor (\ref{eq:SzSz_FF}) is necessary.
With $(1/ \sqrt{N})\sum_{j=1}^Ne^{-\mathrm{i} qj}\left\langle \{\lambda\} | S^z_j  S^z_{j +
1} | \{\mu\} \right\rangle =(1/ \sqrt{N})\sum_{p} e^{\mathrm{i} p} \left\langle \{\lambda\} |S^z_{q-p}S^z_{p}| \{\mu\} \right\rangle$, we give here an explicit expression for the norm squared of theses matrix elements between normalized states
\begin{eqnarray}
  \frac{2\pi}{N}\frac{\left\lvert \sum_{p} e^{\mathrm{i} p}  \left\langle \{\lambda\} |S^z_{q-p}S^z_{p}| \{\mu\} \right\rangle\right\rvert^2}{N (\{\lambda\}) N (\{\mu\})} & = & \frac{N}{16} \prod_i \left| \frac{\phi
  (\lambda_i + \eta / 2)}{\phi (\mu_i + \eta / 2)} \right|^2 \nonumber\\
  & \cdot & \frac{\delta_{q + q_{\lambda} ,q_{\mu}} }{\prod_{\alpha > \beta} \left( \phi^2 (\lambda_{\alpha} -
  \lambda_{\beta}) + \phi^2 (\eta) \right) \vert \Phi ( \{ \lambda \} ) \vert } \nonumber\\
  & \cdot & \frac{\left| \left( e^{i (q_{\lambda} - q_{\mu})} + 1 \right) F_1
  + e^{i (q_{\lambda} - q_{\mu})} F_2 - \left\langle \{\lambda\} | \{\mu\}
  \right\rangle F_3 \right|^2}{ \prod_{j, k = 0}  | \phi (\mu_j - \mu_k + \eta) | | \Phi (\{\mu\}) |}\nonumber \\
  F_1 & = & \det (H(\{\lambda \}, \{\mu \})  - 2 P(\{\lambda \}, \{\mu \}) ) \nonumber\\
  F_2 & = & -4 \prod_i \frac{\phi^{} ( \lambda_i+\eta / 2 )}{\phi^{} ( \mu_i+\eta / 2)} \sum_{n = 1}^M A_n  \left( \det (G^n + B^n) - \det G^n
  \right) \nonumber\\
  F_3 & = &  \prod_{j > k} \phi (\lambda_k - \lambda_j) \frac{\prod_{j < k} \phi (\mu_k - \mu_j) }{\prod_{i } \phi^2 ( \lambda_i+\eta / 2)
  \phi^{- 2} (\mu_i+\eta / 2 )} \nonumber\\
  H_{\tmop{ab}} (\{\lambda \}, \{\mu \}) & = & \frac{\phi (\eta)}{\phi
  (\lambda_a \um \mu_b)} \left( \prod_{k \neq a} \phi (\lambda_k \um
  \mu_b + \eta) - d (\mu_b) \prod_{k \neq a} \phi (\lambda_k \um \mu_b - \eta)
  \right) \nonumber\\
  P_{\tmop{ab}} (\{\lambda_{} \}, \{\mu_{} \}) & = & \frac{\phi (\eta) \prod_k
  \phi (\mu_k \um \mu_b + \eta)}{\phi (\lambda_a \um \eta / 2) \phi (\lambda_a
  + \eta / 2)}
\label{eq:FT_SzSz_FF}
\end{eqnarray}
with $N(\{\lambda\})$ the norm of the state \cite{1981_gaudin_prd_23,1982_korepin_cmp_86,gaudinbook} including the Gaudin determinant $| \Phi (\{\lambda\}) |$. Recall that $ \phi (\lambda)=\lambda$ and $\eta=\mathi$ for the isotropic spin chain.

In the case that the state $\langle\{\lambda\}\rvert$ contains string configurations, columns of the matrix $\Phi$ become equal to leading order in the string deviations $\delta$ and one should use the reduced Gaudin matrix $|\Psi^{(r)}(\{\lambda\})|$ instead (see \cite{kirillov_jms_1988}, \cite{2005_caux_jstat_p09003}).
The other factors in (\ref{eq:FT_SzSz_FF}) are well-defined if only the state $\langle\{\lambda\}|$ contains strings.  However, when the state $\langle\{\mu\}|$ contains string structures, the entries of the matrices $G^n,B^n,H,P$ yield undetermined factors. In case of a single spin form factor, the matrices $H$ and $P$ can be regularized (see \cite{2005_caux_jstat_p09003} and the discussion below \eqref{eq:red_Hmm}). Unfortunately these methods do not apply for the form factor under consideration here. But since the operator $S^z_j S^z_{j+1}$ is Hermitian and the ground state is made of real rapidities, one can always choose $\langle\{\lambda\}|$ to be the state with strings, and obtain well defined results for the form factor.
\subsection{$S^-_{j}S^-_{j+1}$}
The second FF of interest concerns the operator $S^-_j S^-_{j+1}=\sigma^-_j \sigma^-_{j+1}$.
Again, using the result \eqref{eq:ABA_sigma_j} for the local $\sigma^-_j$ operator and simplifying the product of transfer matrices, the form factor reads
\begin{eqnarray}
  \left\langle \{\lambda\}_M \right| \sigma_j^- \sigma_{j + 1}^- \left| \{\mu\}_{M - 2}
  \right\rangle  &=&  \left\langle \{\lambda\}_M \right| \prod_{i = 1}^{j - 1} (A + D) (\xi_i)
  B (\xi_{j }) B (\xi_{j + 1})\prod_{k = j + 2}^N (A + D) (\xi_k) \left|
  \{\mu\}_{M - 2} \right\rangle
\end{eqnarray}
 with the use of the determinant representation and taking the homogeneous limit similarly to (\ref{eq:homo_limit}), the form factor becomes
\begin{eqnarray}
  \left\langle \{\lambda\}_M \right| \sigma_j^- \sigma_{j + 1}^- \left|
  \{\mu\}_{M - 2} \right\rangle & = & \frac{\varphi_{j - 1} (\{\lambda\})}{\varphi_{j - 1} (\{\mu\})}
  \frac{\prod_k^M \phi (\lambda_k + \eta / 2)^2}{\prod^{M - 2}_i \phi (\mu_i +
  \eta / 2)^2} \frac{\det H^{--} (\{\lambda\}_M, \{\mu_i \}_{M -
  2})}{\prod^M_{j > k} \phi (\lambda_k - \lambda_j) \prod^{M - 2}_{j < k} \phi
  (\mu_k - \mu_j)}
\end{eqnarray}

with the $H^{--}$ matrix
\begin{eqnarray}
  H^{--}_{a b} & = & \begin{cases}
    \frac{\phi (\eta)}{\phi (\lambda_a \um \mu_b)} \left( a (\mu_b)
    \prod_{k \neq a} \phi (\lambda_k \um \mu_b + \eta) - d (\mu_b)
    \prod_{k \neq a} \phi (\lambda_k \um \mu_b - \eta) \right) ,\;& b \neq M - 1, M\\
    \frac{\phi (\eta)}{\phi (\lambda_a \um \eta / 2) \phi (\lambda_a + \eta / 2)}
   ,& b = M - 1\\
    \frac{\phi (\eta)\phi(2 \lambda_a)}{\phi^2
    (\lambda_a \um \eta / 2) \phi^2 (\lambda_a + \eta / 2)},& b = M
    \end{cases}	
\label{eq:SmSm_FF}
\end{eqnarray}

\subsubsection{Reduction of determinant for string solutions}
If one considers $\vert \{\mu \} \rangle$ states which include string structures,
following the same procedure as \cite{2005_caux_jstat_p09003}, the determinant representation (\ref{eq:SmSm_FF}) has to be reduced in order to have well-defined matrix elements.
Hereafter, we use two different ways of indexing the rapidities. The notation with one index $\mu_a$ ($a=1,\ldots,M$)
 allows one to identify a row or a column of the matrix under consideration. On the other hand, the notation including three parameters
 $ \mu_\alpha^{k,a}$ ($k=1,\ldots,N_s$, $\alpha=1,\ldots,M_k$ and $a=1,\ldots,n_k$) provides a clear identification of rapidities which belong to the same string. The reduced $H^{--(r)}$ matrix reads then
\begin{eqnarray} 
  &&H^{--(r)}_{a b}  =  \phi (\eta) \frac{F_0 F_1}{J_1} \sum^{n_k}_{i = 0}
  \frac{J_i J_{i + 1}}{F_i F_{i + 1}} \left[ (1-\delta_{i, n} - \delta_{i, 0}) L_{a, i}   -  \left( \delta_{i, n} + \delta_{i, 0} \right) K_{a, i} \right],\quad b:(k,\alpha,1), b\neq M-1,M  \nonumber\\
 && H^{--(r)}_{a b}  =  \phi (\eta) K_{a, i - 1} J_{i - 1},\hspace{5.5cm} b: (k,\alpha,i),\:i = 2, \ldots, n_k
  ,\; b\neq M-1,M \nonumber\\
&&H^{--(r)}_{a (M-1)}=\frac{\phi (\eta)}{\phi (\lambda_a \um \eta / 2) \phi (\lambda_a + \eta / 2)}\; ,\nonumber\\
  &&H^{--(r)}_{a M}=  \frac{\phi (\eta)  \phi (2 \lambda_a )}{\phi^2
    (\lambda_a \um \eta / 2) \phi^2 (\lambda_a + \eta / 2)} \; ,\nonumber\\
  && K_{a, i}  =  \frac{1}{\phi (\lambda_a - \mu_{\alpha}^{k, i})\phi (\lambda_a - \mu_{\alpha}^{k, i+1})}\;,  \hspace{2cm}L_{a, i}  =  - \frac{d}{d \lambda_a}K_{a,i}\; .
\label{eq:red_Hmm}
\end{eqnarray}
\subsubsection{Fourier transform}
In order to express the dynamical structure factor, one needs the Fourier transform of the form factor (\ref{eq:SmSm_FF}) $(1/ \sqrt{N})\sum_{j=1}^Ne^{-\mathrm{i} qj}\left\langle \{\lambda\} | S^-_j  S^-_{j +
1} | \{\mu\} \right\rangle =(1/ \sqrt{N})\sum_{p} e^{\mathrm{i} p} \left\langle \{\lambda\} |S^-_{q-p}S^-_{p}| \{\mu\} \right\rangle$. We express here the norm squared of these matrix elements between normalized states:
\begin{eqnarray} 
 &&\frac{2\pi}{N}\frac{\left\lvert \sum_{p} e^{\mathrm{i} p}  \left\langle \{\lambda\} |S^-_{q-p}S^-_{p}| \{\mu\} \right\rangle\right\rvert^2}{N (\{\lambda\}) N (\{\mu\})}    =  N  \prod_i \left| \frac{\phi (\eta / 2 +
  \lambda_i)^2}{\phi (\eta / 2 + \mu_i)^2} \right|^2 \frac{\delta (q +
  q_{\mu} - q_{\lambda})}{\prod_{\alpha > \beta} \left( \phi^2
  (\lambda_{\alpha} - \lambda_{\beta}) + \phi^2 (\eta) \right) | \Phi (\{\lambda\}) |}\nonumber\\
  && \hspace{4cm}\cdot  \frac{ \left| F \right|^2}{
\left. \prod_{n, m = 0}^{N_s} 
  \prod_{\alpha, \beta = 1}^{M_j,M_k}  \prod^{n_j,n_k}_{i, j = 1, ((n, \alpha, i)
  \neq (m, \beta, j - 1))} | \phi (\mu^j_{\alpha, i} - \mu^k_{\beta,
  j} + \eta) \right|
 | \Phi^{(r)} (\{\mu\}) |}
\label{eq:FT_SmSm_FF}
\end{eqnarray}
with
\begin{eqnarray}
  F & = & \frac{\det H^{--} (\{\lambda\}, \{ \mu \})}{\prod^M_{j > k} \phi (\lambda_k - \lambda_j) \prod^{M - 2}_{j < k} \phi
  (\mu_k - \mu_j)}
\end{eqnarray}
 and with $N (\{\lambda\})$ the norm of the state \cite{1981_gaudin_prd_23,1982_korepin_cmp_86,gaudinbook} including the Gaudin matrix determinant $| \Phi (\{\lambda\}) |$. We recall that $ \phi (\lambda)=\lambda$ and $\eta=\mathi$ for the isotropic spin chain. If the state $\{\mu\}$ contains strings, this results has to be combined with the reduced determinant explicitly given in (\ref{eq:red_Hmm}) and the reduced Gaudin matrix $\Phi^{(r)} (\{\mu\})$ (\cite{kirillov_jms_1988}, \cite{2005_caux_jstat_p09003}).

\section{$S^z_jS^z_{j+1}$ dynamical structure factor in the Heisenberg spin chain}
\label{sec:DSF_S4z}

\begin{figure}[ht]
\centering
\subfigure[]{
\includegraphics[width=0.3\textwidth,angle=270]{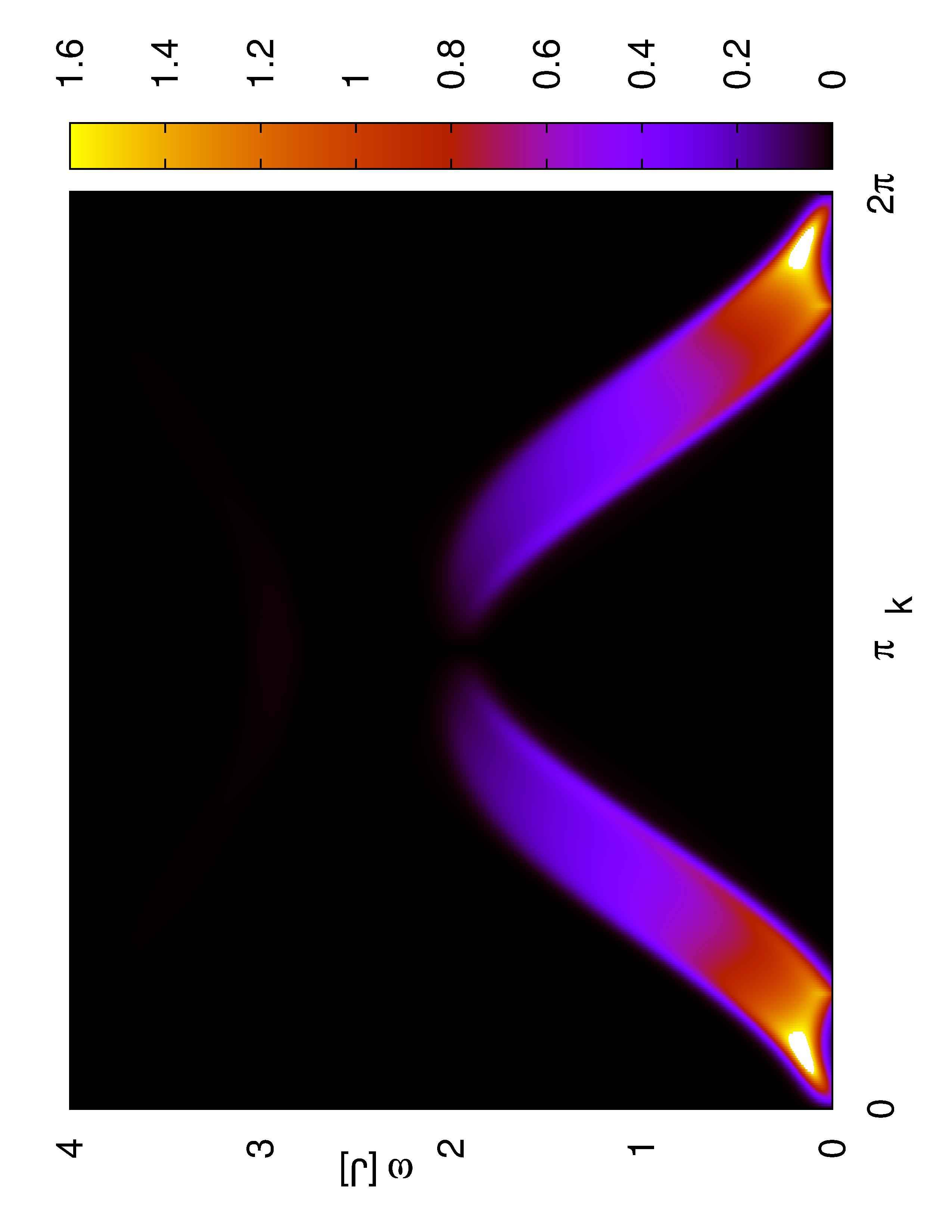}
\label{fig:DSF_SZZ_M50}
}
\subfigure[]{
\includegraphics[width=0.3\textwidth,angle=270]{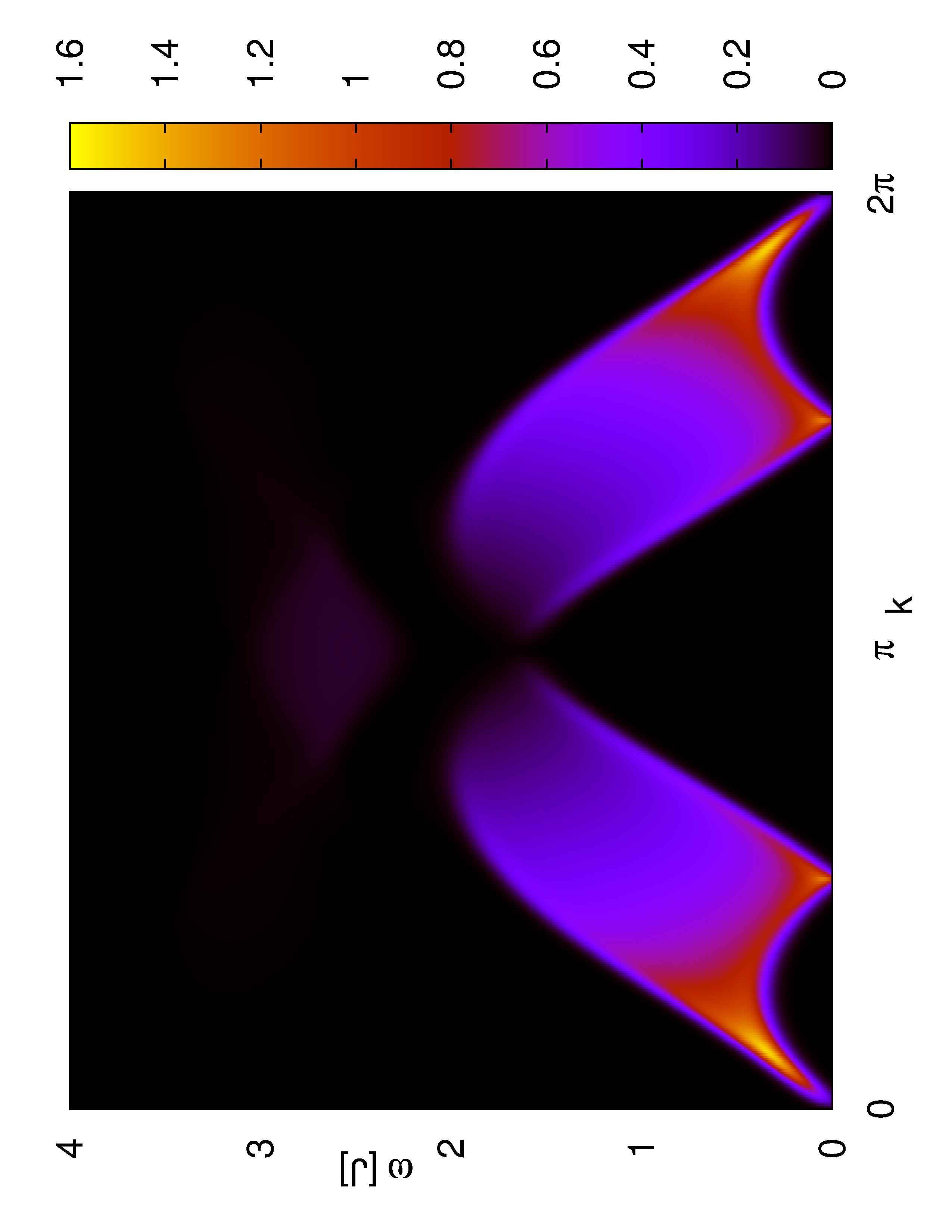}
\label{fig:DSF_SZZ_M100}
}
\subfigure[]{
\includegraphics[width=0.29\textwidth,angle=270]{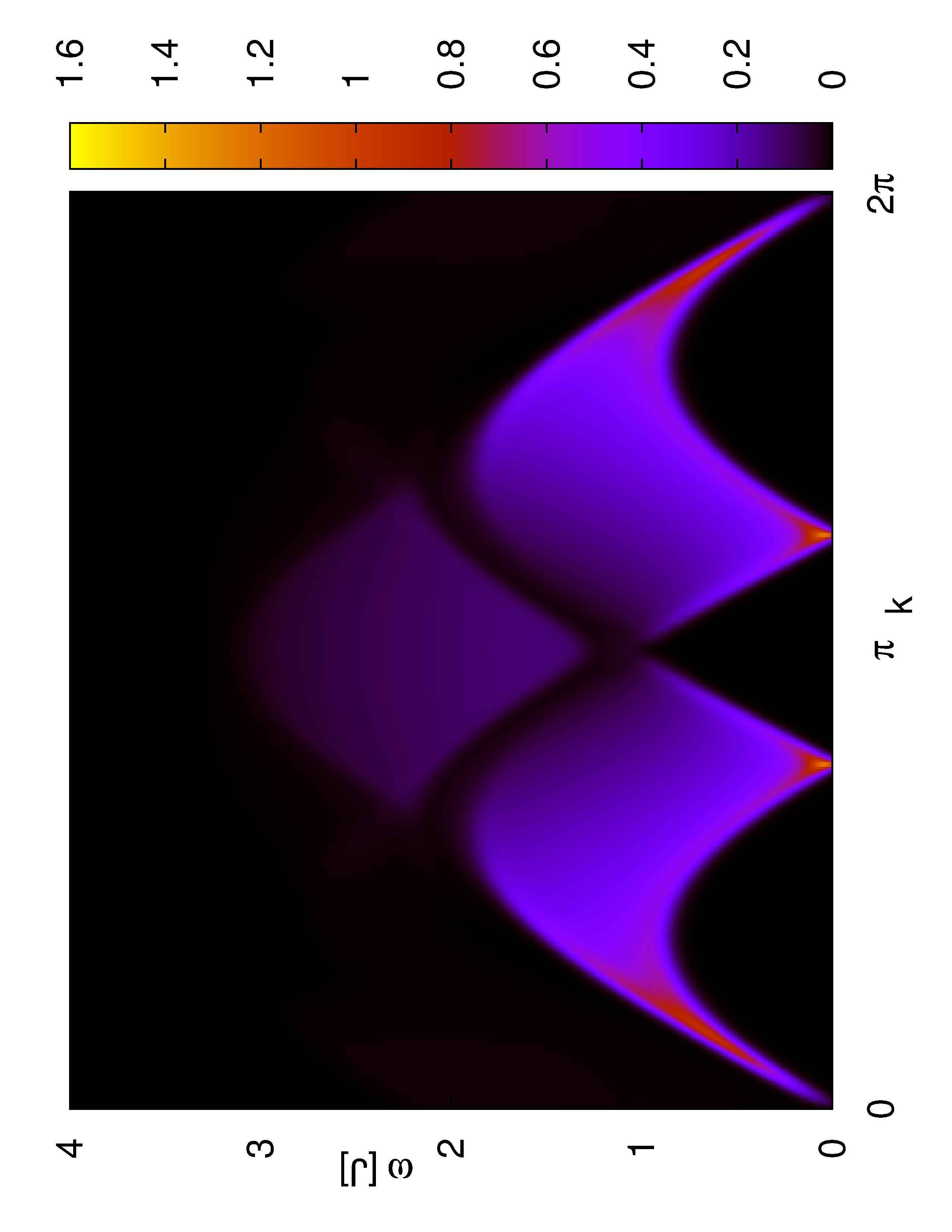}
\label{fig:DSF_SZZ_M150}
}
\subfigure[]{
\includegraphics[width=0.3\textwidth,angle=270]{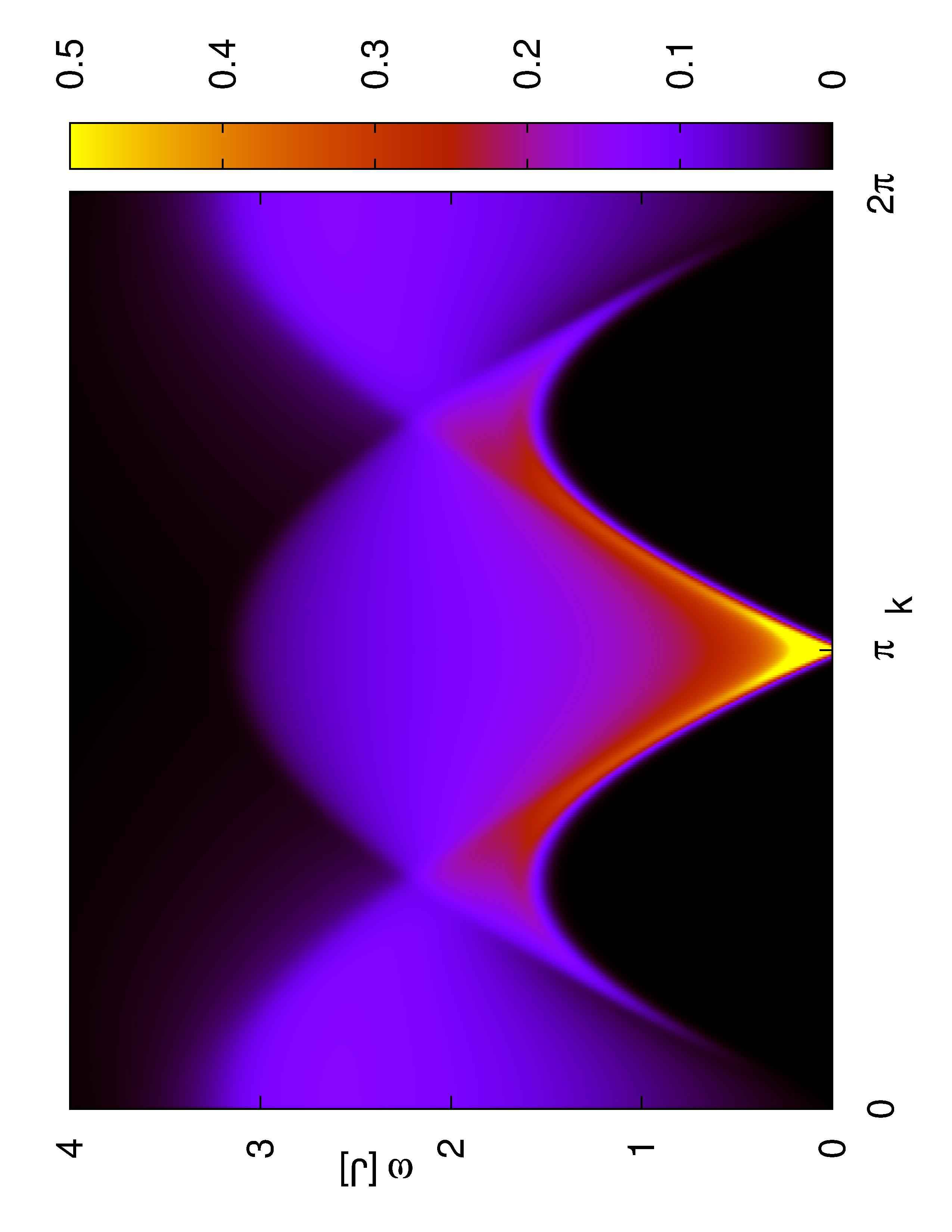}
\label{fig:DSF_SZZ_M200}
}
\caption{$S_c^{4z}(q,\omega)$ DSF for a $N=400$ XXX spin chain and \subref{fig:DSF_SZZ_M50} $M=50$,
 \subref{fig:DSF_SZZ_M100} $100$, \subref{fig:DSF_SZZ_M150} $150$ and \subref{fig:DSF_SZZ_M200} $200$}
\label{fig:DSF_SZZ}
\end{figure}

In order to have the response of the system to an excitation of momentum $q$ and energy $\omega$, we express the dynamical structure factor \eqref{eq:adjacent_DSF} in Lehmann representation with the Fourier transform form factor:
\begin{eqnarray}
S^{4z}(q,\omega)&=&\frac{2 \pi}{N}\sum_{\alpha} \Big|\sum_{p} e^{\mathrm{i} p}  \left\langle GS \right| S^z_{q-p}S^z_{p} \left| \alpha \right \rangle  \Big|^2\delta(\omega-\omega_\alpha)
\label{eq:DSF_SzSz}
\end{eqnarray}
where $ \langle GS | S^z_{q-p}S^z_{p}| \alpha \rangle$ is the form factor between the normalized ground state and the whole set of normalized eigenstates
 and $\omega_{\alpha } = E_\alpha-E_{GS}$ is the relative energy.
 We will explain hereafter that in order to cover all the contributing eigenstates $| \alpha \rangle $,
 we express the results in terms of the $S^z_{j}S^z_{j+1}$ and $S^-_{j}S^-_{j+1}$ form factors.

\subsection{Spin sectors}
\label{sec:SzSz_sectors}
As the $S^z_{q-p}S^z_{p}$ conserves the spin $S^{tot}_z$ and if we consider the form factor with a ground state with $M$ reversed spins,
 the matrix elements are only non-zero for eigenstate with the same magnetization.
To access these states in the different sectors but with $S_z^{tot}=N/2-M$, we include the eigenstates that contain one and two infinite rapidities,
 and with the contributions of all sectors, the DSF reads:
 $S^{4z}(q,\omega)=\sum^{M}_{M'=0}\sum_{\alpha_{M'} } \Big| \sum_{p} e^{\mathrm{i} p} \left\langle GS_M \right| S^z_{q-p}S^z_{p}(S^-_{0})^{(M-M')} \left| \alpha_{M'} \right \rangle  \Big|^2$ 
with $\left\langle GS_M \right|$ the normalized ground state of magnetization $S^{tot}_z=\frac{N}{2}-M$ and $\left| \alpha_{M'} \right \rangle$ a normalized eigenstate including $M'$ rapidities.
As the ground state is also highest-weight, $\sum_i S_i^+ |GS_M\rangle=0 $ and by commutation we identify the only three contributing sectors
\begin{eqnarray}
&& \hspace{-2cm}\sum_{\alpha} \Big|\sum_{p} e^{\mathrm{i} p}  \left\langle GS_M \right| S^z_{q-p}S^z_{p} \left| \alpha \right \rangle  \Big|^2 =\sum_{ \alpha_{M}} \Big|\sum_{p} e^{\mathrm{i} p} \left\langle GS_M \right| S^z_{q-p}S^z_{p} \left| \alpha_{M}  \right \rangle  \Big|^2 \nonumber\\
&&\hspace{2cm}+\frac{1}{N-2M+2} \sum_{\alpha_{M-1}} \Big|\sum_{p} e^{\mathrm{i} p}  \left\langle GS_M \right| S^-_{q-p}S^z_{p}+S^z_{q-p}S^-_{p} \left| \alpha_{M-1} \right \rangle  \Big|^2 \nonumber\\
&&\hspace{2cm}+ \frac{2}{(N-2M+3)(N-2M+4)}\sum_{\alpha_{M-2}} \Big| \sum_{p} e^{\mathrm{i} p} \left\langle GS_M\right| S^-_{q-p}S^-_{p} \left| \alpha_{M-2}  \right \rangle  \Big|^2 
\label{eq:sectors}
\end{eqnarray}
where the first and last Fourier transform form factors given in (\ref{eq:FT_SzSz_FF}) and (\ref{eq:FT_SmSm_FF}).
In the case of zero magnetic field, ($M=N/2)$, the contribution of the sector $M-1$ is identically zero.
 Indeed, following the Wigner-Eckart theorem \cite{sakurai_book}, the matrix elements can be expressed proportionally to a Clebsch-Gordan coefficient.
We decompose the indexing of the Bethe states $\left| \alpha \right \rangle$ into a state parameter, total spin and spin-$z$ number:
 $\left| \beta, S^{tot}, S_z^{tot} \right \rangle$ and we rewrite the form factor as:
\begin{eqnarray}
\left\langle GS \right| S^z_{j}S^z_{j+1} \left| \alpha \right \rangle & \propto & \sum_{\beta,s^{tot},s^{tot}_z}\left\langle GS,0,0 \right| S^z_{j}
 \left| \beta,s^{tot},s^{tot}_z \right \rangle \left \langle \beta,s^{tot},s^{tot}_z \right|S^z_{j+1}\left| \alpha,S^{tot},0 \right \rangle\nonumber\\
& \propto & \sum_{s^{tot},s^{tot}_z} \left\langle 0,1;0,0 | s^{tot},s^{tot}_z\right \rangle \left \langle s^{tot},1;s^{tot}_z,0 |S^{tot},0 \right \rangle\nonumber\\
& \propto & \left\langle 0,1;0,0 | 1,0\right \rangle \left \langle 1,1;0,0 |S^{tot},0 \right \rangle
\end{eqnarray}
with $\left \langle j_1,j_2;m_1,m_2 |J,M \right \rangle$ the Clebsch-Gordan coefficient.
Then the only non-zero contribution sectors are $S^{tot}=\frac{N}{2}-M,\frac{N}{2}-M+2$.
 
One notices from \eqref{eq:sectors} that for large spin chain, in a finite magnetic field,
the contributions of the sectors $M-1$ and $M-2$ are in order of $\frac{1}{N}$ and $\frac{1}{N^2}$, so negligible.
In summary, to evaluate the dynamical structure factor $S^{4z}(q,\omega)$ one needs the $S^z_{q-p}S^z_{p}$ form factor in sector $S^{tot}=\frac{N}{2}-M$
and for a magnetization close to zero ($N-2M \sim 1$), one must add the contributions of $S^-_{q-p}S^-_{p}$ form factor in sector $S^{tot}=2$.
\subsection{Sum rules}
The sum rules are of particular importance for the calculation since they provide checks independent of the algebraic Bethe Ansatz.
The integrated intensities allow one to control the efficiency of the algorithm by summing all the computed form factors.
\subsubsection{$S^z_jS^z_{j+1}$ integrated intensity} 
\label{sec:SzSz_int_int}
\begin{eqnarray}
  \frac{1}{N} \sum_q \frac{1}{2 \pi} \int d \omega S^{4 z} (q, \omega) & = &
  \frac{1}{N^2} \sum_q \sum_{j, j'} e^{- \mathi q (j - j')} \left\langle
  \tmop{GS} \right| S^z_j S^z_{j + 1} S^z_{j'} S^z_{j' + 1} \left| \tmop{GS}
  \right\rangle
   =  \frac{1}{16} \; .
\end{eqnarray}
This result is valid for every value of the magnetic field and for any anisotropy.

\subsubsection{$S^z_jS^z_{j+1}$ first frequency moment} 
\label{sec:SzSz_ffm}
\begin{eqnarray}
  \frac{1}{2 \pi} \int d \omega \omega S^{4 z} (q, \omega)
 & = & \frac{1}{N}  \sum_{\mu}\sum_{j,j'} e^{-
  \mathi q (j-j')} \left\langle \tmop{GS} \right| H  S^z_j S^z_{j + 1} - S^z_j S^z_{j + 1} H \left| \mu
  \right\rangle \left\langle \mu \right|  S^z_{j'} S^z_{j' + 1}
  \left| \tmop{GS} \right\rangle\nonumber\\
  & = & \frac{J}{N 2}  \sum_{j, j'} e^{- \mathi q (j - j')} \left\langle \tmop{GS} \right| \left[ \left[ \sum_i S^x_i S^x_{i
  + 1} + S^y_i S^y_{i + 1}, S^z_j S^z_{j + 1} \right], S^z_{j'} S^z_{j' + 1}
  \right] \left| \tmop{GS} \right\rangle\nonumber\\
  & =& \frac{J}{ N} \sum_j   \left\langle\tmop{GS} \right|\frac{1}{ 4} \left(S^x_j S^x_{j +1}
 + S^y_j S^y_{j + 1} \right) - \cos (2 q) S^z_{j - 1} \left(
S^x_j S^x_{j + 1} + S^y_j S^y_{j + 1} \right) S^z_{j+ 2} \left| \tmop{GS} \right\rangle \; .
\end{eqnarray}
For $h=0$ and in the isotropic spin chain, we use the expectation values calculated in \cite{2003_sakai_pre_67} to evaluate the first frequency moment that reads then
\begin{eqnarray}
J \left[ \frac{\ln (2)}{6} -\frac{1}{24}  + \cos (2 q) \left( \frac{1}{120} -
\frac{\ln 2}{2} + \frac{169}{120} \zeta (3) - \frac{5}{6} \zeta (3) \ln 2 -
\frac{3}{10} \zeta (3)^2 - \frac{65}{48} \zeta (5) + \frac{5}{3} \zeta (5) \ln
2 \right)  \right] \; .
\end{eqnarray}

\section{$S^-_jS^-_{j+1}$ dynamical structure factor in the Heisenberg spin chain}
\label{sec:DSF_Smm}

\begin{figure}[h]
\centering
\subfigure[]{
\includegraphics[width=0.3\textwidth,angle=270]{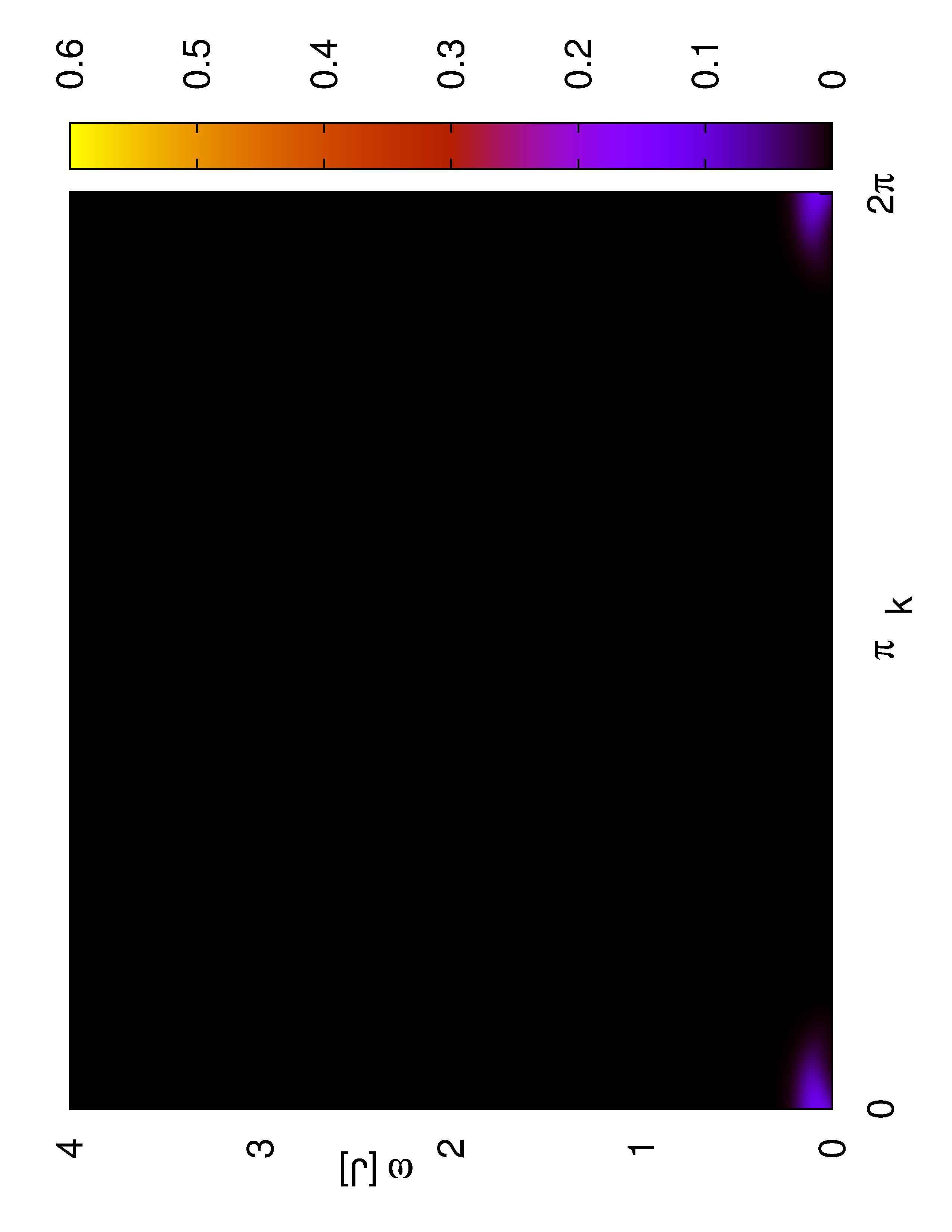}
\label{fig:DSF_Smm_M50}
}
\subfigure[]{
\includegraphics[width=0.3\textwidth,angle=270]{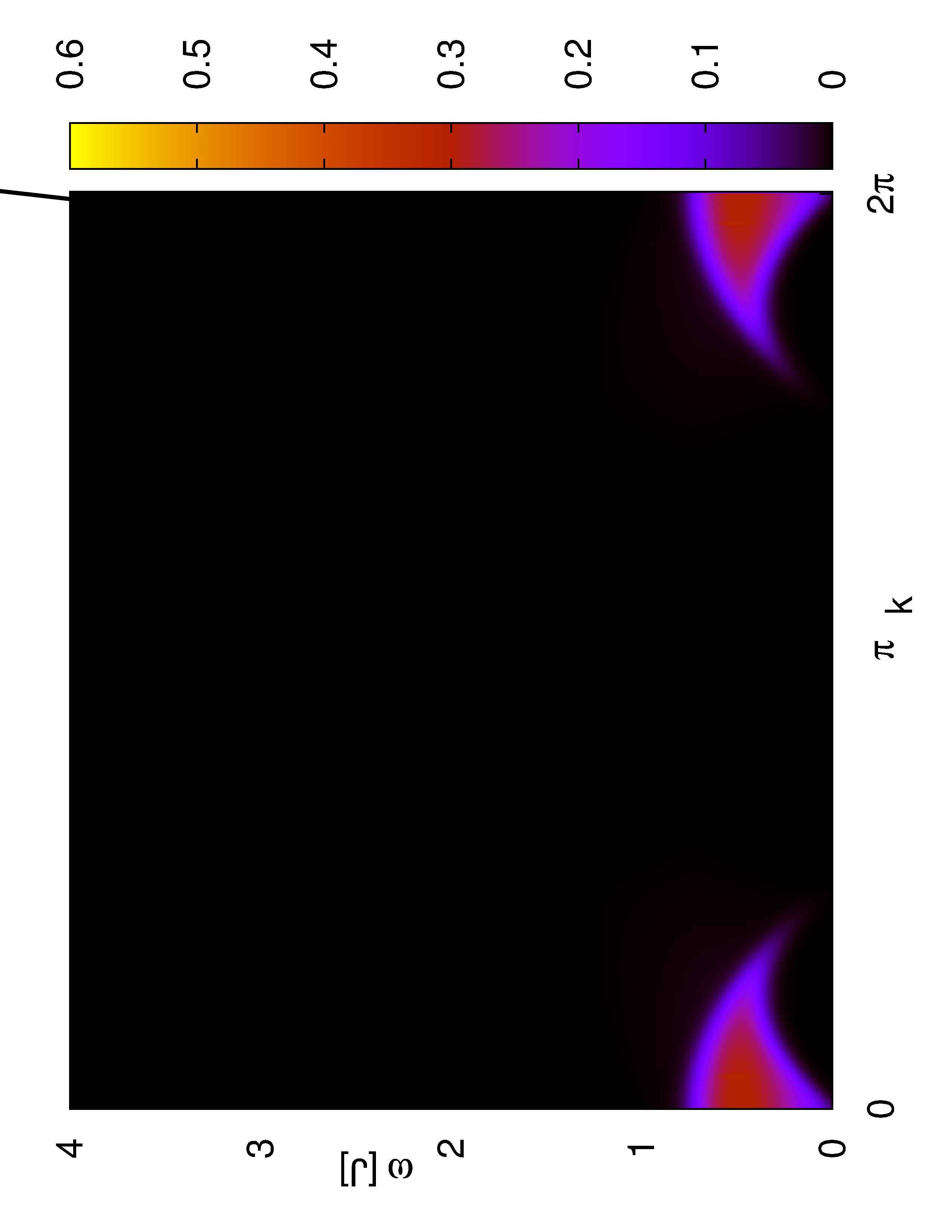}
\label{fig:DSF_Smm_M100}
}
\subfigure[]{
\includegraphics[width=0.3\textwidth,angle=270]{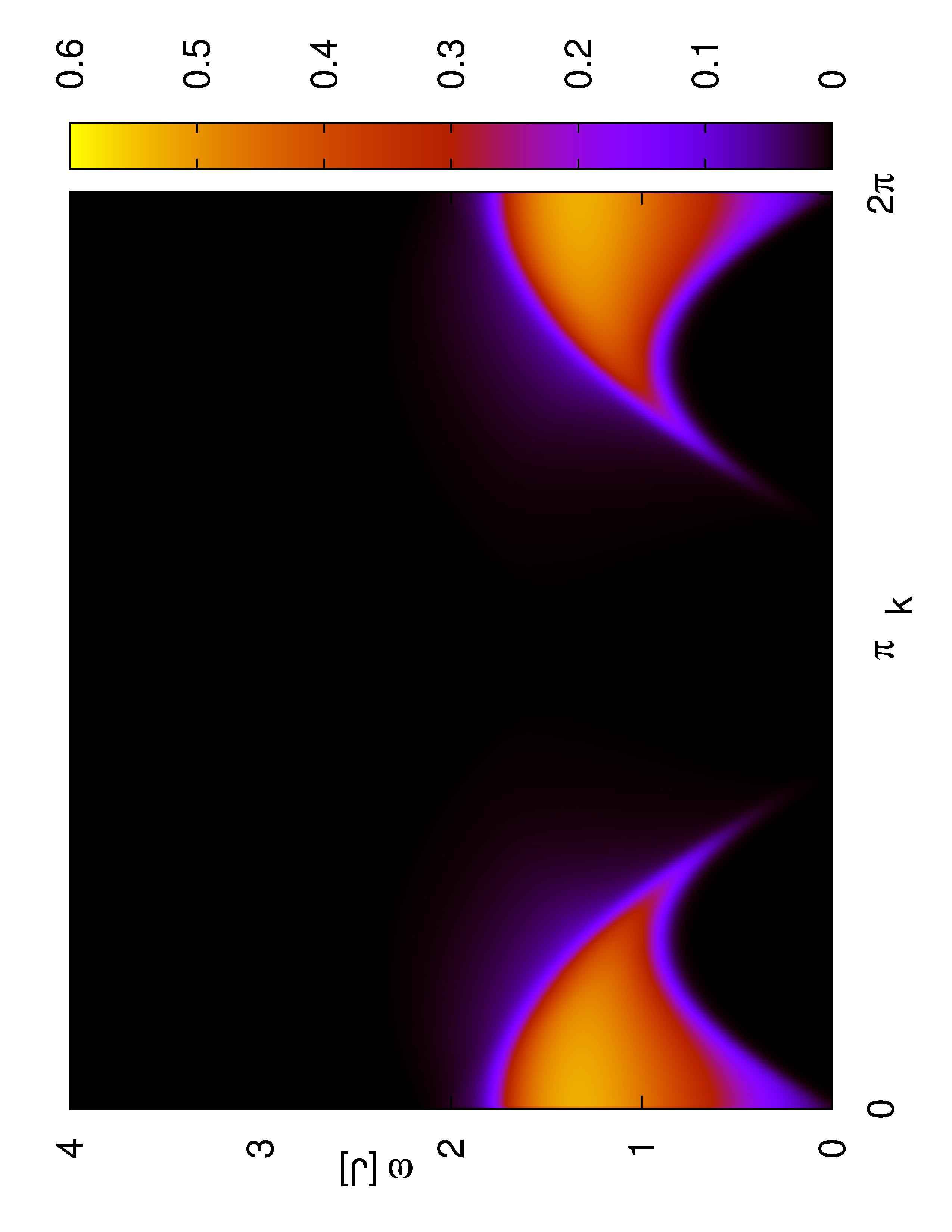}
\label{fig:DSF_Smm_M150}
}
\subfigure[]{
\includegraphics[width=0.3\textwidth,angle=270]{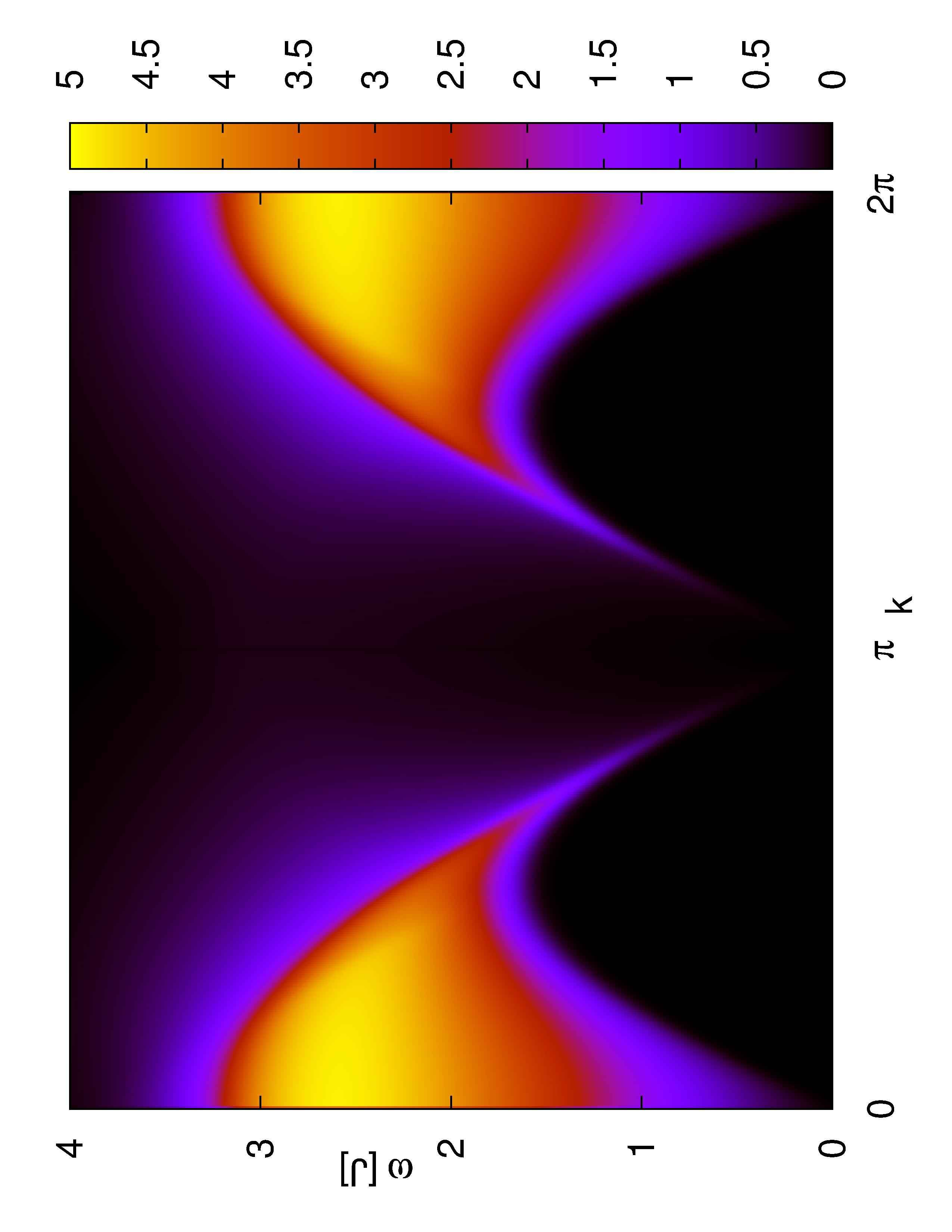}
\label{fig:DSF_Smm_M200}
}
\caption{$S^{--++}(q,\omega)$ DSF in a $N=400$ XXX spin chain   and \subref{fig:DSF_Smm_M50} $M=50$,
 \subref{fig:DSF_Smm_M100} $100$,  \subref{fig:DSF_Smm_M150} $150$ and  \subref{fig:DSF_Smm_M200} $200$.  }
\label{fig:DSF_Smm}
\end{figure}

For two adjacent local spin raising operators, using the result (\ref{eq:FT_SmSm_FF}), the dynamical structure factor reads 
\begin{eqnarray}
S^{--++}(q,\omega)&=&\frac{2 \pi}{N}\sum_{\alpha} \Big| \sum_{p} e^{\mathrm{i} p} \left\langle GS \right| S^-_{q-p}S^-_{p} \left| \alpha \right \rangle  \Big|^2\delta(\omega-\omega_\alpha)
\label{eq:DSF_SmSm}
\end{eqnarray}
where $ \langle GS | S^-_{q-p}S^-_{p}| \alpha \rangle$ is the form factor between the normalized ground state and the whole set of normalized eigenstates
 and $\omega_{\alpha } = E_\alpha-E_{GS}$ is the relative energy.
The operator $ S^m_{q-p}S^m_{p}$ flipping down two spins and by conservation of the $z$-spin value,
 no sector of $M'>M-2$ is contributing to the dynamical structure factor. Moreover following the same procedure as (\ref{eq:sectors}) by acting with an $S^-_{q=0}$
 operator, it is straightforward to show that the contributions of higher spin sector for $M'<M-2$ are all null and that if the ground state belongs to the $M$th sector,
 the only non-zero form factor are with eigenstate of sector $M-2$, $| \alpha_{M-2} \rangle $.
  \subsection{Sum rules}	
In this subsection, the integrated intensity and the first frequency moment of the $S^-_jS^-_{j+1}$ operator are calculated. 
\subsubsection{$S^-_jS^-_{j+1}$ integrated intensity} 
Summing all the dynamical structure factor over all frequency and momenta gives the intergrated intensity:
  \label{sec:SmSm_int_int}
  \begin{eqnarray}
  \frac{1}{N} \sum_q \frac{1}{2 \pi} \int d \omega S^{\um -} (q, \omega) 
  & = & \frac{1}{N} \sum_j \left\langle \tmop{GS} \right| S^-_j S^+_j S^-_{j
  + 1} S^+_{j + 1} \left| \tmop{GS} \right\rangle \nonumber\\
  & = & \frac{1}{4} - \left( \frac{1}{2} - \frac{M}{N} \right) + \left( \frac{1}{N}
\frac{\partial E_0}{\partial \Delta} + \frac{1}{4} \right)
\end{eqnarray}
where we used the Hellmann-Feynman theorem for the evaluation of $\sum_j\left \langle \text{GS} \right \rvert S^z_j S^z_{j+1} \left\lvert \text{GS} \right\rangle$. The result is valid for $h\geq0$ and for the isotropic as well as for the anisotropic spin chain.  

\subsubsection{$S^-_jS^-_{j+1}$ first frequency moment at zero magnetic field} 
\label{sec:SmSm_ffm}
The results hereafter are restricted to the case of zero magnetic field
 since we use several rotational symmetries in order to keep the result concise.
\begin{eqnarray}
\frac{1}{2 \pi} \int d \omega \omega S^{- -} (q, \omega)
 & = & \frac{1}{2 N}  \sum_{j, j'} e^{- \mathi q (j - j')} \left\langle \tmop{GS} \right|
\left[ \left[ H, S^-_j S^-_{j + 1} \right], S^+_{j'} S^+_{j' + 1} \right]
\left| \tmop{GS} \right\rangle\nonumber\\
& = &
  \frac{J}{2 N} \sum_j - 4 \cos (2 q) \left\langle S^z_j S^z_{j + 1} S^z_{j +
  2} S^z_{j + 3} \right\rangle
  + 2 \cos (2 q) \left\langle S^+_j S^-_{j + 1} S^z_{j + 2} S^z_{j + 3}
  \right\rangle\nonumber\\
  &-& (6 \cos (2 q) + 4 \cos (q)) \left\langle S^+_j S^z_{j + 1} S^z_{j + 2}
  S^-_{j + 3} \right\rangle
  + (4 \cos (q) - 2 \cos (2 q)) \left\langle S^+_j S^z_{j + 1} S^-_{j + 2}
  S^z_{j + 3} \right\rangle\nonumber\\
  &+& (2 - 2 \cos (q)) \left\langle S^z_j S^z_{j + 2} \right\rangle
  + (4 + 2 \cos (q)) \left\langle S^z_j S_{j + 1}^z \right\rangle \; .
\end{eqnarray}
We denote here $\left \langle O  \right\rangle$ the expectation value of the operator $O$ at zero temperature. 
This result is only valid for isotropic spin chain.

If we use the results of \cite{2003_sakai_pre_67} to evaluate the correlators, the first frequency moment becomes
\begin{eqnarray}
\frac{1}{2 \pi J} \int d \omega \omega S^{- -} (q, \omega)
 & = &  2 \ln (2)-\frac{3}{4} \zeta (3)  - \frac{1}{4} + \cos (q) \left[ \zeta (3) -  \frac{4}{3} \ln (2) \right ] \nonumber\\
& +& \cos (2 q) \left[ \ln (2) \left( 10 \zeta (5) - \frac{8}{3} - 5 \zeta (3) 
\right) - \frac{65}{8} \zeta (5) + \zeta (3) \left(
\frac{41}{5} - \frac{9}{5} \zeta (3) \right) + \frac{1}{20} \right] \; .
\end{eqnarray}

\section{Identification of the excitations in the isotropic spin chain}
\label{sec:catalog}
\begin{figure}[ht]
\centering
\includegraphics[width=0.8\textwidth,angle=0]{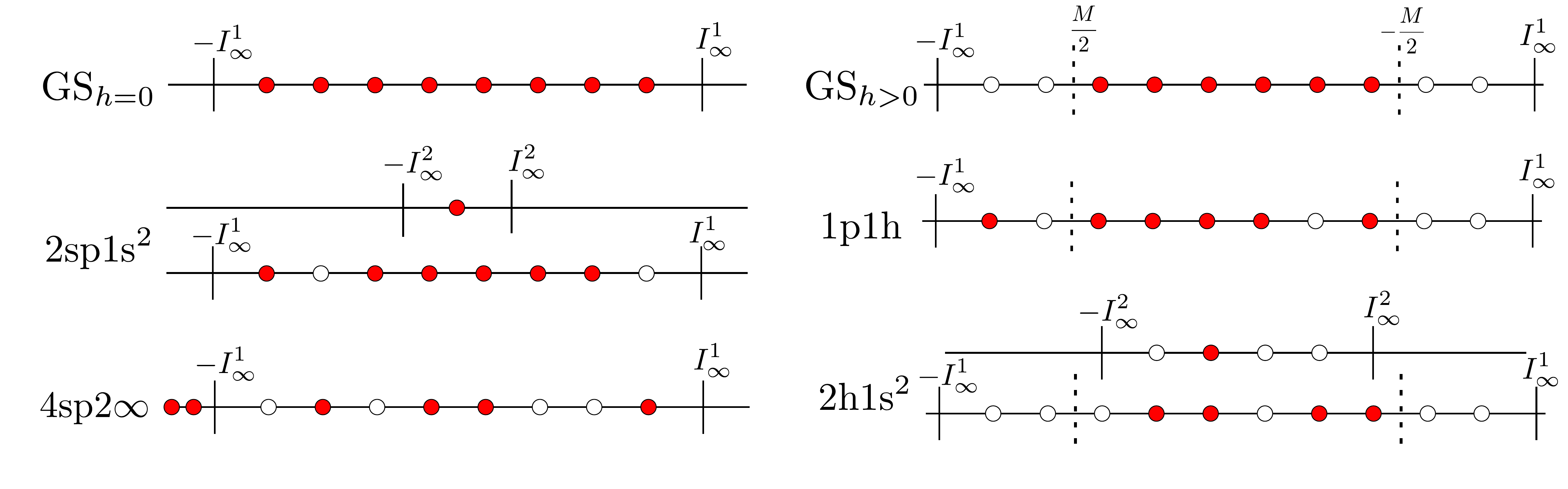}
\label{fig:classification}
\caption{Combinations of quantum numbers corresponding to the ground state and several excited states (see text for description) in the $N=16$ spin chain.
}
\end{figure}

We give in this section a description of the excitations in terms of quasiparticles that occur in the Heisenberg spin chain. This allows one to identify the contributions to each correlations. 
We divide the description of the excited states into two parts: zero and non-zero magnetic field,
 since the polarization modifies drastically the combinatorics of the states above the ground state.
The identification of the states is made with the use of the set $\{ I^j_{\alpha} \}$ (see (\ref{eq:Bethe}),(\ref{eq:Bethe-Takahashi_eq}))
and the excitations of the spin chain are constructed from all the possible combinations of quantum numbers within the 
 boundaries \cite{1971_takahashi_ptp}
\begin{equation}
 \lvert I^j_\alpha\rvert < I^j_\infty =  \frac{1}{2} \left[ N +1 -\sum_{k=1}^{N_s} M_k \left( 2 \min(n_j,n_k) -\delta_{j,k}\right) \right] \; .
\end{equation}
\subsection{Excitations at $h=0$}
In the case of zero polarization, the ground state quantum numbers occupy all the possible vacancies for finite real rapidities,
 i.e. $\{ I^1_\alpha=\alpha -\frac{M+1}{2} \},\> \alpha=1,\ldots,M$.
 The creation of excitations requires then to either take a rapidity to infinity, create a string or to remove rapidities.
 \subsubsection{String excitations}
The creation of an $n$ string configuration allows vacancies in the quantum numbers for real rapidities.
The string state so created is non-dispersive since there is only one accessible quantum number,
but the holes created in the set $\{I^1_\alpha\}$, denoted  ``spinon'' quasiparticles \cite{1981_faddeev_pla_85}, have a continuum spectrum of excitations.
The presence of the complex rapidities imposes via the Bethe equations the number of spinons created and for each $n$ string, there are $2n-2$ spinons in the excited state. 
As notation, we use, e.g.  $ 2\text{sp} 1\text{s}^2$ for a state with $2$ spinons and one $2$ string.

 \subsubsection{Spin raising and infinite rapidities}

There are two other processes which create spinons by removing quantum numbers from the ground state.

First, if one acts with a local spin raising operator, i.e.: $S^+_j$, the spin chain will change sector from $M=N/2$ to $M'=N/2-1$ in an excited state with $S^{tot}=S_z^{tot}=1$. This is equivalent to removing one number from the set $\{I^1_\alpha\}$.
The quantum numbers go then from integers to half integers or vice versa, and consequently each spin raising creates two spinons.
 We denote this excitation with $2\text{sp},4\text{sp},\ldots$ for two, four and higher number of spinons.

Second, the Bethe equations allow one to send rapidities to infinity and therefore to remove the corresponding quantum numbers.
 Similarly to the spin raising operation, this creates a pair of spinons for each rapidity sent to infinity but on the contrary of the spin lowering operation,
 the excited has $S_z^\text{tot}=0$ but $S^\text{tot}=1$. 
The infinite rapidity is actually equivalent to a global spin raising operation:$ \left \lvert \{\lambda,\infty\}_M \right \rangle = \frac{1}{\sqrt{N}} \sum_j S^-_j \left \lvert \{\lambda\}_{M-1} \right \rangle$
  where $ \left \lvert \{\lambda,\infty\}_M \right \rangle$ belongs then to a higher spin sector but with the same $S^\text{tot}_z$.
 These states are described in detail in section  \ref{sec:spin_setup} and they contribute to other spin sectors (see section \ref{sec:SzSz_sectors}). The notation we use is 
$2\text{sp}{\infty},4\text{sp}{2\infty},\ldots$ for excitations of one, two, etc. number of infinite rapidities.

For the three different kinds of excitations described above, the quasiparticles created, spinons,
 have always the same dispersion relation and therefore the spectra are identical even though the states are actually different.
\subsection{Excitations at $h>0$}
In presence of a magnetic field, the ground state of the spin chain is polarized, $S^\text{tot}_z > 0$, and
 therefore there are $N-2M$ vacancies in the quantum numbers.
Similarly to the $h=0$ case, the lowest energy state forms a Fermi like sea in the quantum numbers:
 $\{ I^1_\alpha=\alpha -\frac{M+1}{2} \},\> \alpha=1,\ldots,M$, although there are vacancies between the highest/lowest quantum numbers and the boundaries:$\frac{M-1}{2}+1<I^j_\infty$.
We discuss here three excitations that can occur above the ground state.

 \subsubsection{Particle-hole}
Due to the presence of vacancies, a quantum numbers from the ground state can be taken to higher value. One creates this way a pair of quasiparticles: a hole left in the Fermi sea and a particle evolving above the Fermi surface.
We write in this article these excited states by $i\text{p}i\text{h}$ for $i$ particle-hole pairs.

 \subsubsection{Spin raising}
Similarly to the $h=0$ case, acting with a spin raising operator $S^+_j$ on the spin chain,
removes one rapidity and creates a hole in the Fermi sea. The change in qantum numbers parity change also the number of vacancies above the Fermi sea
but we only count as quasiparticle the holes below the Fermi surface. Therefore, applying $i$ spin raising operators on the ground state
 creates an excited state with $i$ holes belonging to the sector $M'=M-i$ and which we denote $i\text{h}$.
In finite magnetic field, the creation of infinite rapidities is still admissible by the Bethe equations and provides similar hole quasiparticles
 but as discussed in \ref{sec:SzSz_sectors}, the form factors of such states decay like the inverse of the spin chain length.
 
\subsubsection{String excitations}
Excited states can be created by forming a string configuration in the rapidities. In contrast to $h=0$,
 a $n$ string leaves $n$ holes in Fermi sea of rapidities. Moreover,
 the boundary $I^n_\infty$ is of ${\mathcal O}(N)$ and then much bigger than at $h=0$.
 Therefore the string states have dynamics that contribute to the spectrum. The notation we use for a $i$ string state is 
$ i \text{h} 1 \text{s}^i$.
\section{DSF Evaluations}
\label{sec:DSF_eval}
In this section we discuss the evaluation of the dynamical structure factors (DSF) $S^{4z}(q,\omega)$ and $S^{--++}(q,\omega)$ computed following the schemes \ref{sec:DSF_S4z} and \ref{sec:DSF_Smm} and the contributions of the different kinds of excitations are analyzed in the principal cases.

Considering a spin chain of length $N=400$ and with fillings of $M=200,150,100,50$,
the value of the DSF is computed by summing over the eigenstate contributions.
The summation over the intermediate states is performed to obtain quantitative results over the Brillouin zone and over an energy range that covers all the significant weight. This is done using the ABACUS algorithm \cite{2009_caux_jmp_50} which sums intermediate state contributions in a close to optimal order. 
From a given set of quantum numbers, we solve the Bethe equations (\ref{eq:Bethe}), and calculate the energy and momentum (\ref{eq:E&P}). The value of the form factor is then computed with the formula (\ref{eq:FT_SzSz_FF},\ref{eq:FT_SmSm_FF}). 

The saturation of the integrated intensity is checked along with the first frequency moment in order to control the completion of the computation. From a general point of view, the saturation could be improved by either lengthening the time of computation or improving the scan efficiency. 
To obtain smooth curves in frequency $\omega$, the delta function in (\ref{eq:DSF_SzSz},\ref{eq:DSF_SmSm}) is broadened in energy to a scale commensurate with the level spacing. 
\subsection{$S^{--++}(q,\omega)$} 

\begin{table}[h]
\centering
$S^{--++}(q,\omega)$
\\
\begin{tabular}{c c c c | c c}
( \% ) & $M=50$ & $M=100$ & $M=150$& & $M=200$\\ 
\hline \hline
 $2\text{h}$ & 98.1 & 88.3 & 64.3&  $4\text{sp}$& 90.8\\
 $1\text{p}3\text{h}$ & 1.9 & 11.2 & 29.8& $6\text{sp}1\text{s}^2$  &1.4\\
$2\text{p}4\text{h}$ & 0.01 & 0.4 & 2.4\\
\hline
$\sum_q\int d\omega S^{--++}$ & 99.9 & 99.8  & 96.6  &   & 92.2\\
\hline	
\hline
\end{tabular}
\caption{Contributions of excitation types to the the $S^{--++}(q,\omega)$ DSF for $N=400$ and $M=50,100,150$ and $200$. The last row gives the saturation of the integrated intensity .}
\label{tab:SmSm}
\end{table}

The $S^{--++}(q,\omega)$ evaluation is shown in figure \ref{fig:DSF_Smm} for the four different magnetizations.
 One observes that the shape is similar to the single spin raising $S^{-+}(q,\omega)$ DSF \cite{2009_kohno_prl_102}
 but with a shift of $\pi$ in momentum because the two spin raising operator removes two quantum numbers to the ground state instead of one. There is also a significant difference in the distribution of the signal. For the two spin raising DSF,
 the weight is more homogeneously spread from low to high energy although for a single spin raising, it is mainly located at the lower boundary of the spectrum.
$S^{--++}(q,\omega)$ is the propagator of a pair of neighboring spin up and
 as a consequence, it vanishes when the polarization of the spin chain tends to saturation ($M=0$). In table \ref{tab:SmSm},
one can observe also that as the magnetic field increases,
 the number of excitations with particle-hole pairs created by the double spin raising decreases. In figure \ref{fig:Excit_type_DSF_Smm}, we split the total signal of $S^{--++}(q,\omega)$ into the three main contributing types of excitations. The signal becomes weaker as the number of particles in the excited state increases and it is also noticeable that if the $2$h type carries weight at low energy, the signal of the $1$p$3$h and $2$p$4$h vanishes at zero energy and is mainly significant at higher energy.
Besides, the DSF at $h=0$ ($M=200$) shows a signal which is mainly composed of $4$-spinon excitations (table \ref{tab:SmSm}). We divide the $S^{--++}(q,\omega)$ DSF signal into the $4$-spinon and $6$-spinon excitations in figure \ref{fig:Excit_type_DSF_Smm}. In addition to the fact that the $6$sp$1$s$^2$ contribution is globally two orders of magnitude lower than for the $4$sp, we notice that the $6$sp$1$s$^2$ weight is essentially inexistent between $q=\pi/2$ and $q=3\pi/2$.

\begin{figure}[h]
\centering
\begin{tabular}{ccc}

\subfigure[$\;2$h]{
\includegraphics[width=0.3\textwidth]{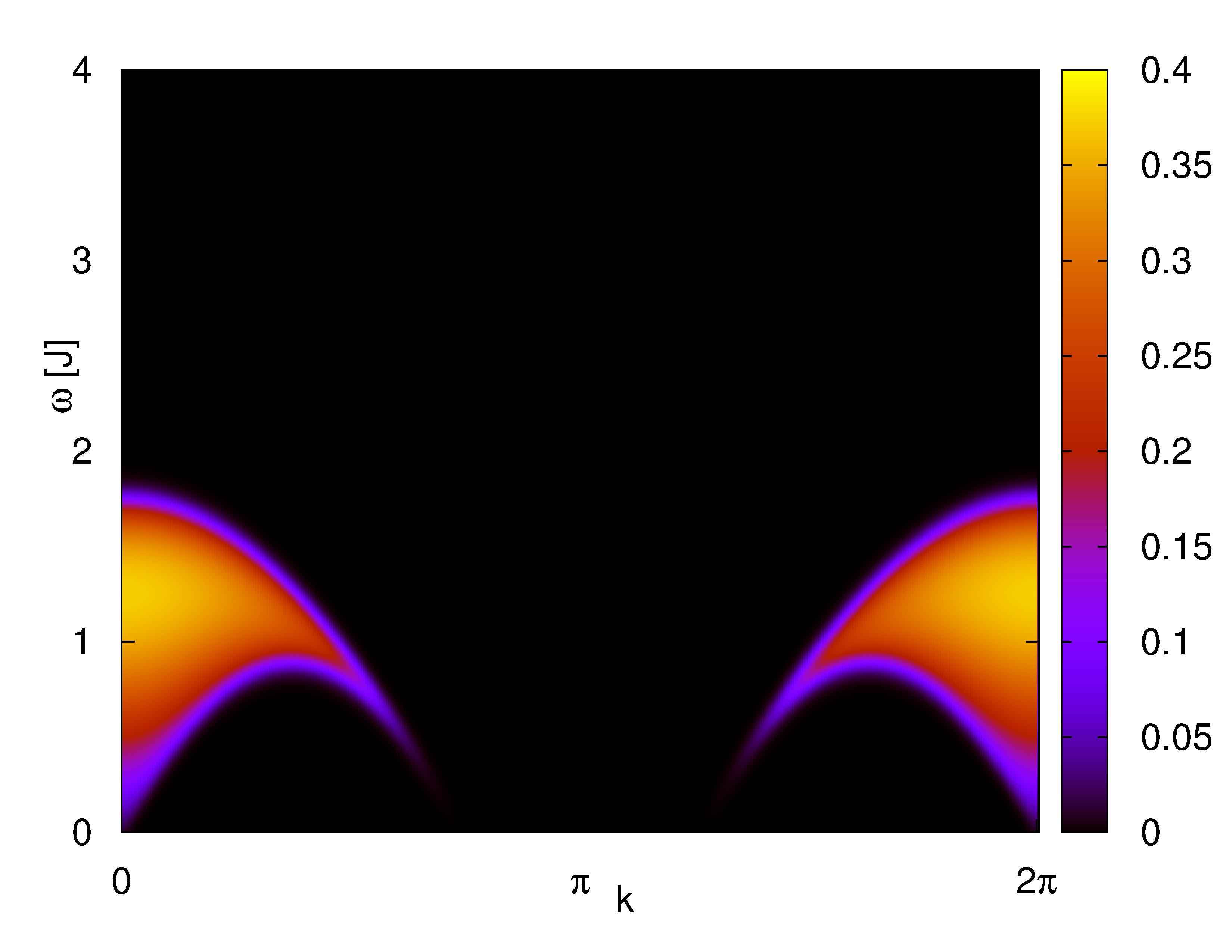}
\label{fig:0p_DSF_Smm_M150}
}
&\subfigure[$\;1$p$2$h]{
\includegraphics[width=0.3\textwidth]{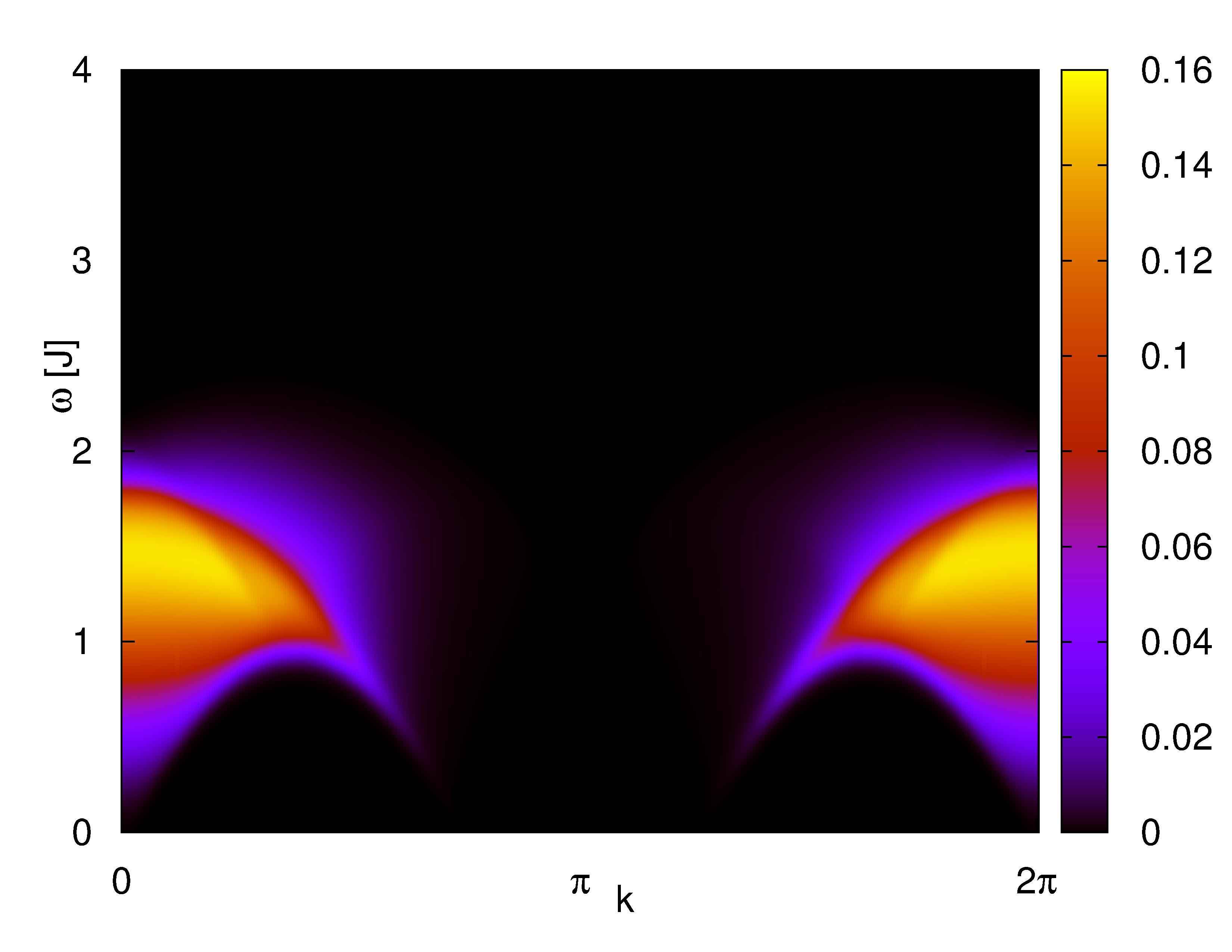}
\label{fig:1p_DSF_Smm_M150}
}
&\subfigure[$\;2$p$2$h]{
\includegraphics[width=0.3\textwidth]{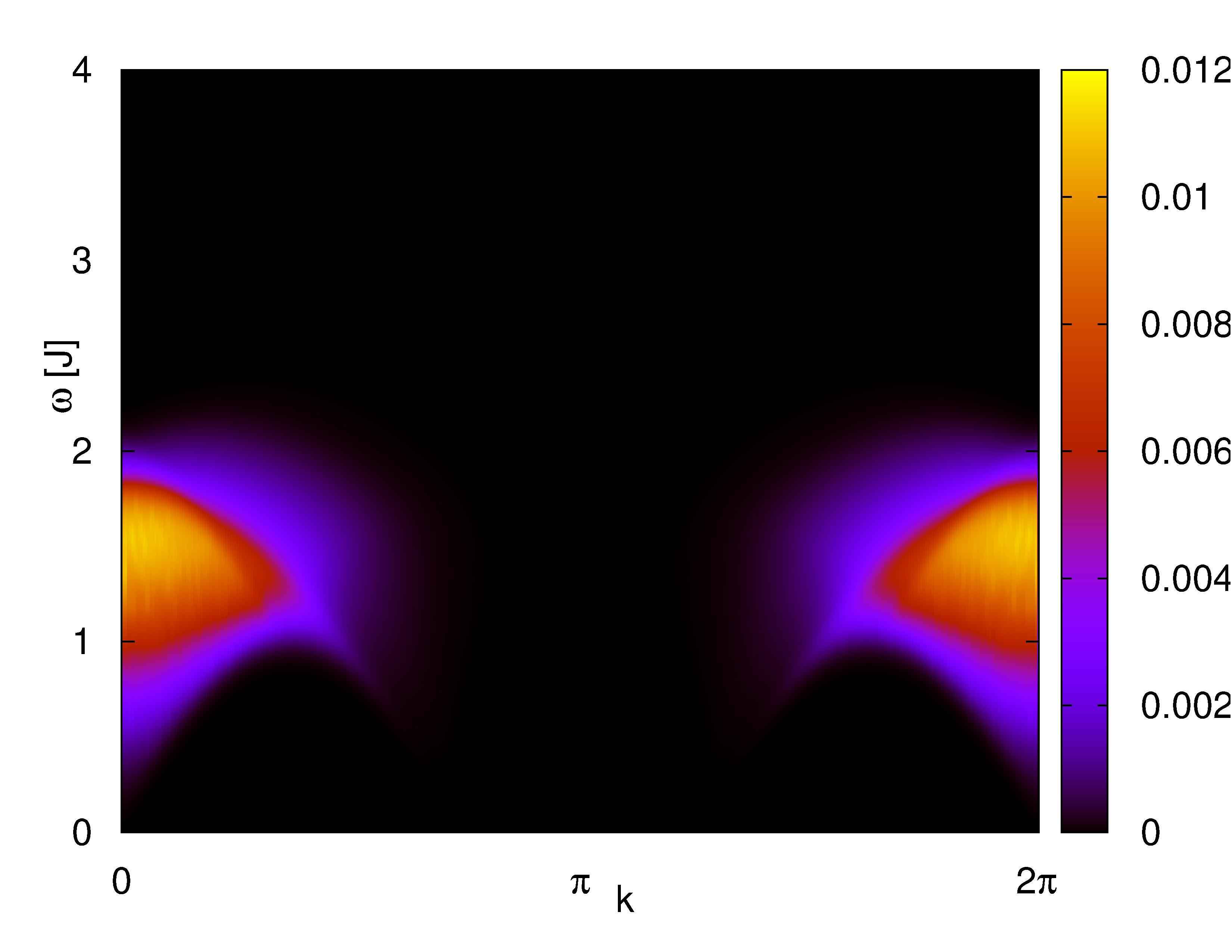}
\label{fig:2p_DSF_Smm_M150}
}
\end{tabular}
\begin{tabular}{cc}
\subfigure[$\;4$sp]{
\includegraphics[width=0.3\textwidth]{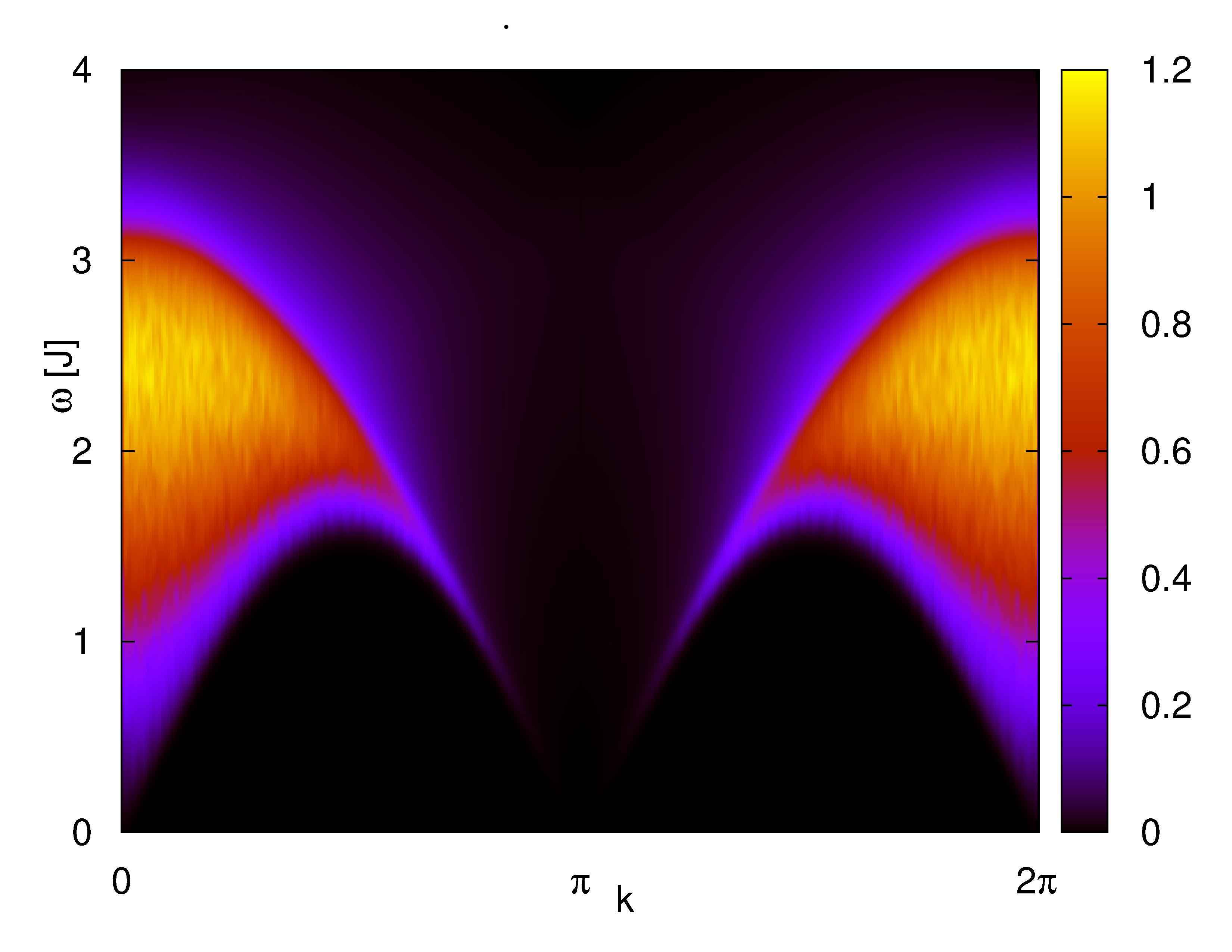}
\label{fig:4sp_DSF_Smm_M200}
}
&\subfigure[$\;6$sp$1$s$^2$]{
\includegraphics[width=0.3\textwidth]{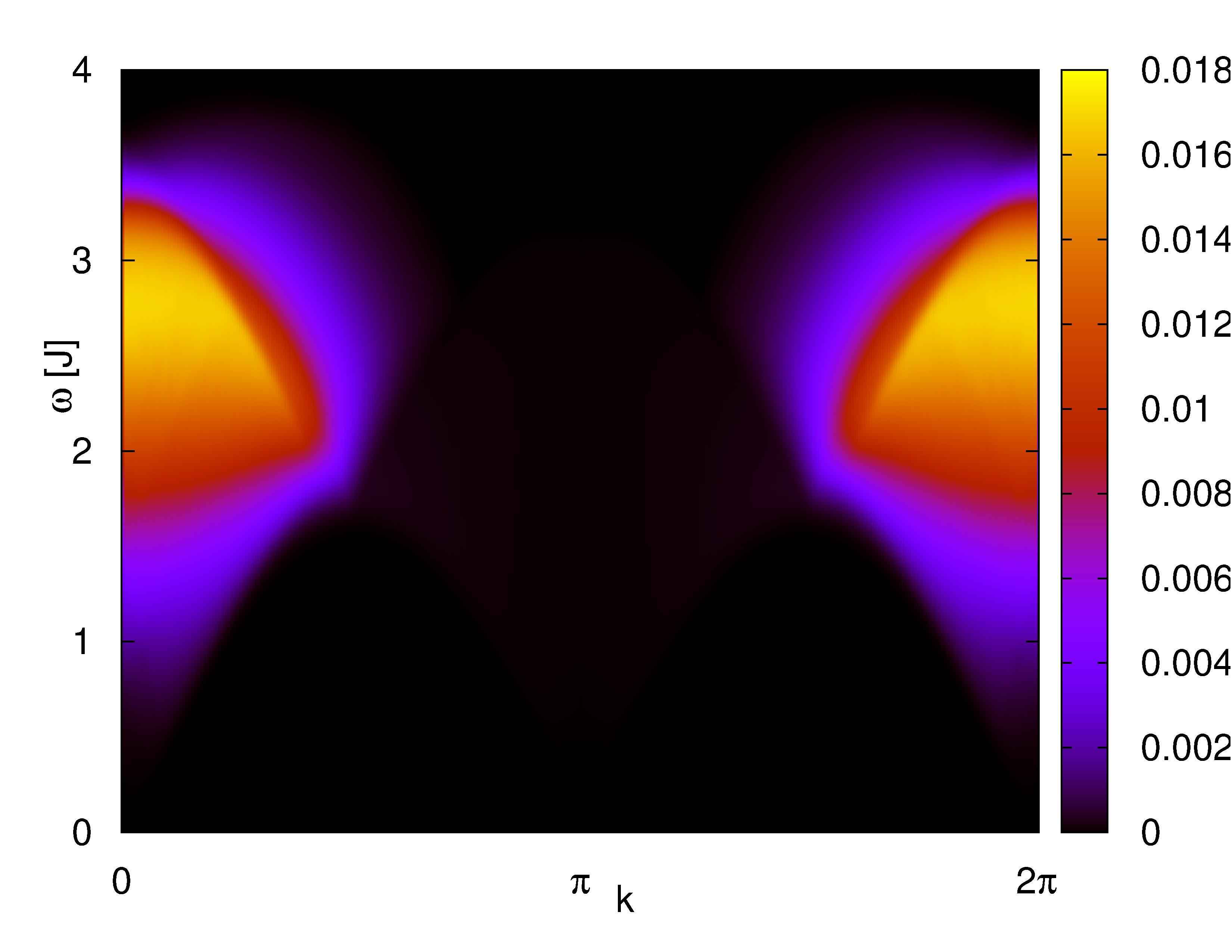}
\label{fig:6sp1s_DSF_Smm_M200}
}
\end{tabular}

\caption{Contributions by type of excitation to the $S^{--++}(q,\omega)$ DSF in a $N=400$ spin chain for $M=150$ in \subref{fig:0p_DSF_Smm_M150}, \subref{fig:1p_DSF_Smm_M150} and \subref{fig:2p_DSF_Smm_M150} or for $M=200$ in \subref{fig:4sp_DSF_Smm_M200} and \subref{fig:6sp1s_DSF_Smm_M200}.  The graphs are labeled by the type of excitation that are represented.}
\label{fig:Excit_type_DSF_Smm}
\end{figure}

\subsection{$S^{4z}(q,\omega)$}

The $S^{4z}(q,\omega)$ DSF is here numerically evaluated and the signal weight is analyzed and attributed to the different kinds of excitations. In order to emphasize the dynamic properties, we consider only the connected DSF defined as 
$S_c^{4z}(q,\omega)=S^{4z}(q,\omega)$, $\forall(q,\omega)\neq(0,0)$
and $S_c^{4z}(q=0,\omega=0)=0 $.
\begin{table}[h]
\centering
$S_c^{4z}(q,\omega)$
\\
\begin{tabular}{c c c c| c c}
( \% ) & $M=50$ & $M=100$ & $M=150$& & $M=200$\\ 
\hline \hline
 $1\text{p}1\text{h}$ & 93.2 & 86.4 & 64.8&$2\text{sp}1\text{s}^2$ & 57.4 \\
 $2\text{p}2\text{h}$ & 5.3 & 7.0 & 14.4&$4\text{sp}{2\infty}$ & 38.0\\
 $3\text{p}3\text{h}$ & 0.1 & 0.1 & 0.2 &$6\text{sp}{2\infty}1\text{s}^2$ & 0.6\\
 $2\text{h}1\text{s}^2$ & 0.1 & 1.2 & 9.1 &$4\text{sp}2\text{s}^2$ & 0.003 \\
\hline
$\sum_q\int d\omega S_c^{4z}$ & 98.68 & 94.69  & 89.01  &   & 96.0\\
\hline
\hline
\end{tabular}
\caption{Percentages of contributions of excitation types to the the $S_c^{4z}(q,\omega)$ DSF for $N=400$ and $M=50,100,150$ and $200$. The last row gives the saturation of the integrated intensity.}
\label{tab:SzSz}
\end{table}

As pictured in figure \ref{fig:DSF_SZZ}, in the presence of a magnetic field the signal
 is very similar to the single spin $S^{zz}(q,\omega)$ DSF \cite{2005_caux_jstat_p09003, 2009_kohno_prl_102} and as shown in table \ref{tab:SzSz},
  the contribution of the $2$ string states are of the same order. From the table \ref{tab:SzSz}, one notices also that the percentage of the $1$ and $2$ particle-hole excitations to the sum rule is approximately stable as the magnetic field decreases. From the analysis of the computation results, we can identify where the different kinds of excited eigenstates supply weight and we show in figures \ref{fig:1p_DSF_SZZ_M150}, \ref{fig:2p_DSF_SZZ_M150} and  \ref{fig:1s_DSF_SZZ_M150} the contributions of each type of excitation. For the case $M=150$, we split the  $S^{4z}_c(q,\omega)$ DSF into the three main weight carrying types. The $1$p$1$h states contribute mainly at low energy whereas the $2$p$2$h states carry their highest signal in the upper part of the DSF map. In addition, the weight of eigenstates with a single $2$ string structure is located around $q=\pi$ and $\omega=2$ and this category of excitation is gapped in energy. Although the string is accompanied by $2$ holes,
the signal clearly shows the presence of this bound state which is difficult to observe \cite{2011_ganahl_arxiv}.

The signal of the $S^{4z}_c(q,\omega)$ DSF at zero magnetic field takes the original form shown in figure \ref{fig:DSF_SZZ}.
 As shown in table \ref{tab:SzSz}, most of the weight carried by the spin singlet eigenstates ($S_{tot}=0$) corresponds to $2$sp$1$s$^2$ excitations. By definition, these excited states are located inside the $2$-spinon spectrum, i.e.: $\frac{\pi}{2} \lvert\sin(q)\rvert <\omega_{2\text{-spinon}} <\pi \sin(q/2)$ \cite{1981_faddeev_pla_85}. Almost all the rest of the weight is carried by $4$-spinons states ($6$ and higher spinons states are negligible) and falls within $4$-spinon spectrum: $\frac{\pi}{2} \lvert\sin(q)\rvert <\omega_{4\text{-spinon}} < \text{max}\{2\pi \sin(q/4),2\pi \sin(q/4+\pi/2) \}$. We see on table \ref{tab:SzSz} that among the $4$-spinon states, the spin quintuplet eigenstates (with two infinite rapidities) contribute the most to the signal and singlet states e.g. $4$sp$2$s$^2$ are insignificant.
We divide the $S^{4z}_c(q,\omega)$ DSF into its main contributing eigenstates in figures \ref{fig:2sp_DSF_SZZ_M200} and \ref{fig:4sp2inf_DSF_SZZ_M200}. These illustrations show clearly the difference in the momentum and energy localization of the $2$-spinon and $4$-spinon signal. Moreover, although the weight distribution of the $4$sp$2\infty$ is clearly broader in energy, the $2$sp$1$s$^2$ signal is highly concentrated around $q=\pi$ and zero energy. The remarkable contribution of the $4$-spinon clearly differentiates with the $S^{zz}(q,\omega)$ DSF at $h=0$ \cite{2009_kohno_prl_102, 2006_caux_jstat_p12013}.


\begin{figure}[h]
\centering
\begin{tabular}{ccc}
\subfigure[ $\;1$p$1$h]{
\includegraphics[width=0.3\textwidth]{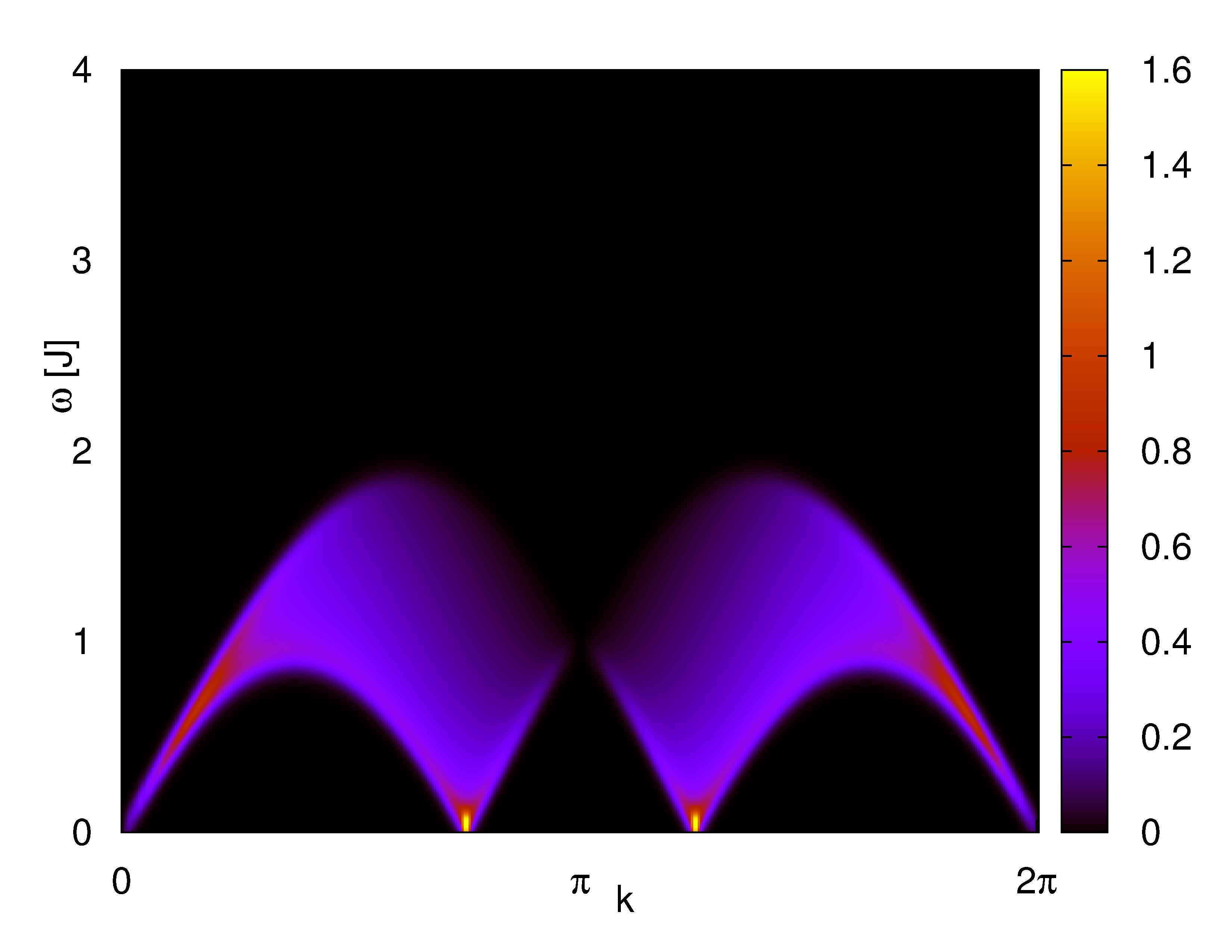}
\label{fig:1p_DSF_SZZ_M150}
}
&\subfigure[ $\;2$p$2$h]{
\includegraphics[width=0.3\textwidth]{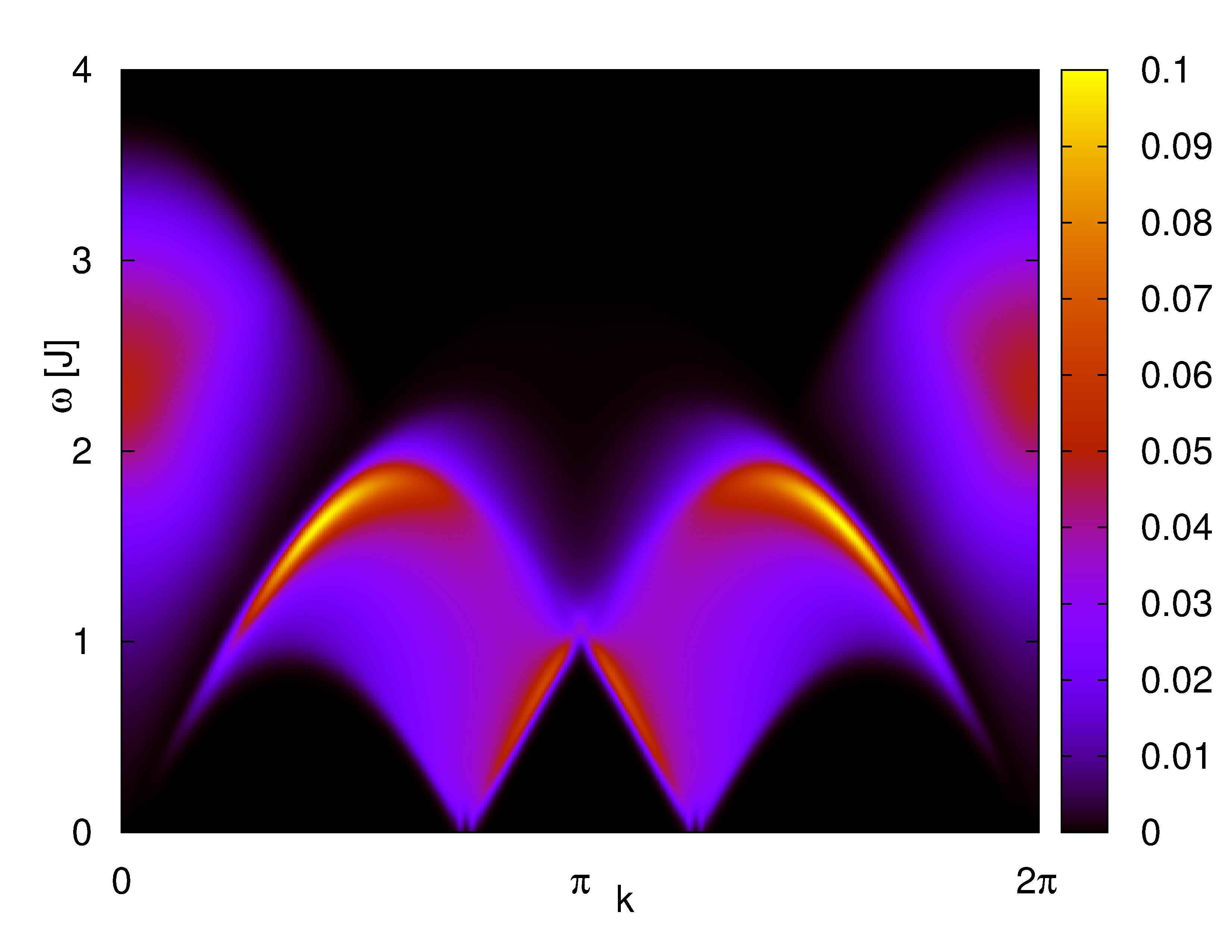}
\label{fig:2p_DSF_SZZ_M150}
}
&\subfigure[ $\;2$h$1$s$^2$]{
\includegraphics[width=0.3\textwidth]{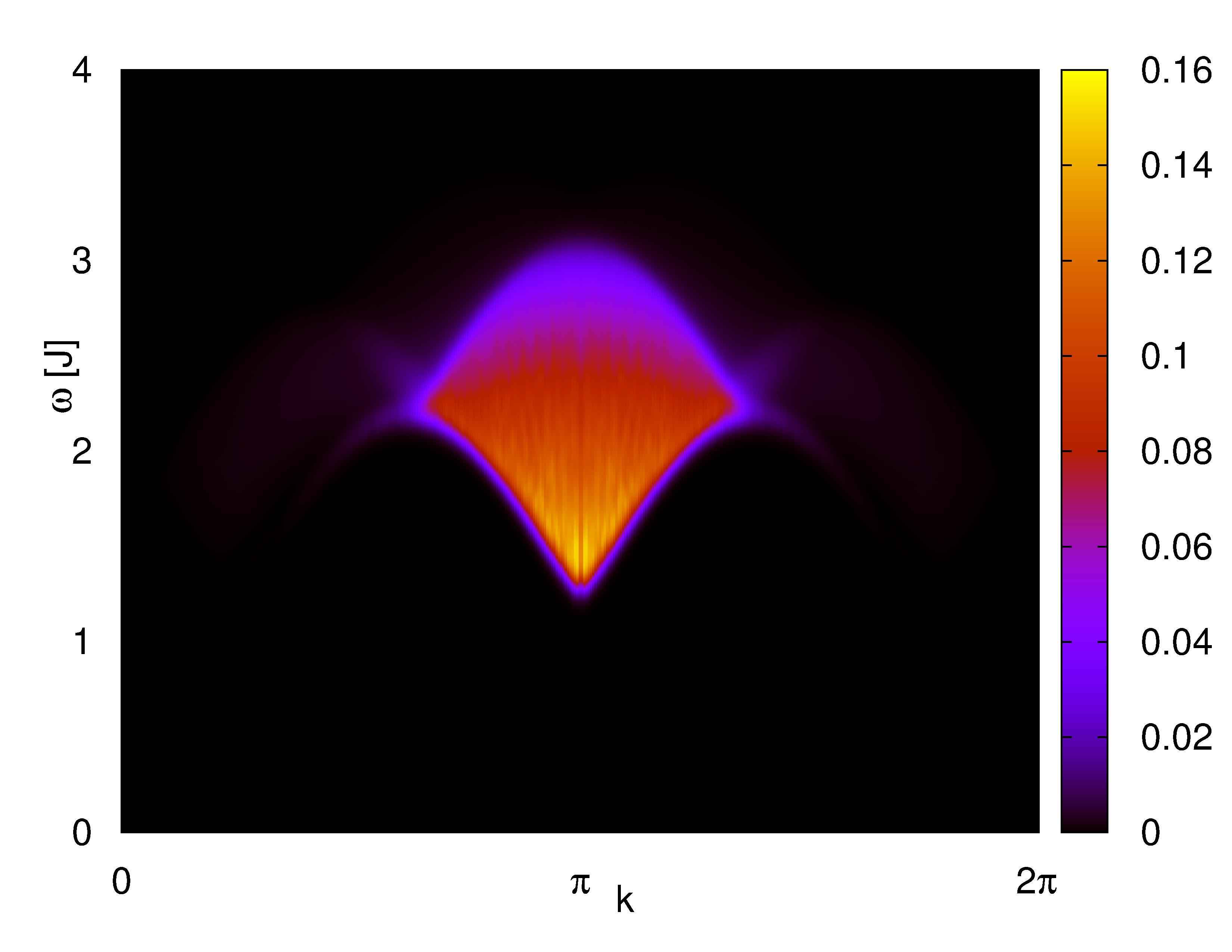}
\label{fig:1s_DSF_SZZ_M150}
}
\end{tabular}

\begin{tabular}{cc} 
\subfigure[$\;2$sp$1$s$^2$]{
\includegraphics[width=0.3\textwidth]{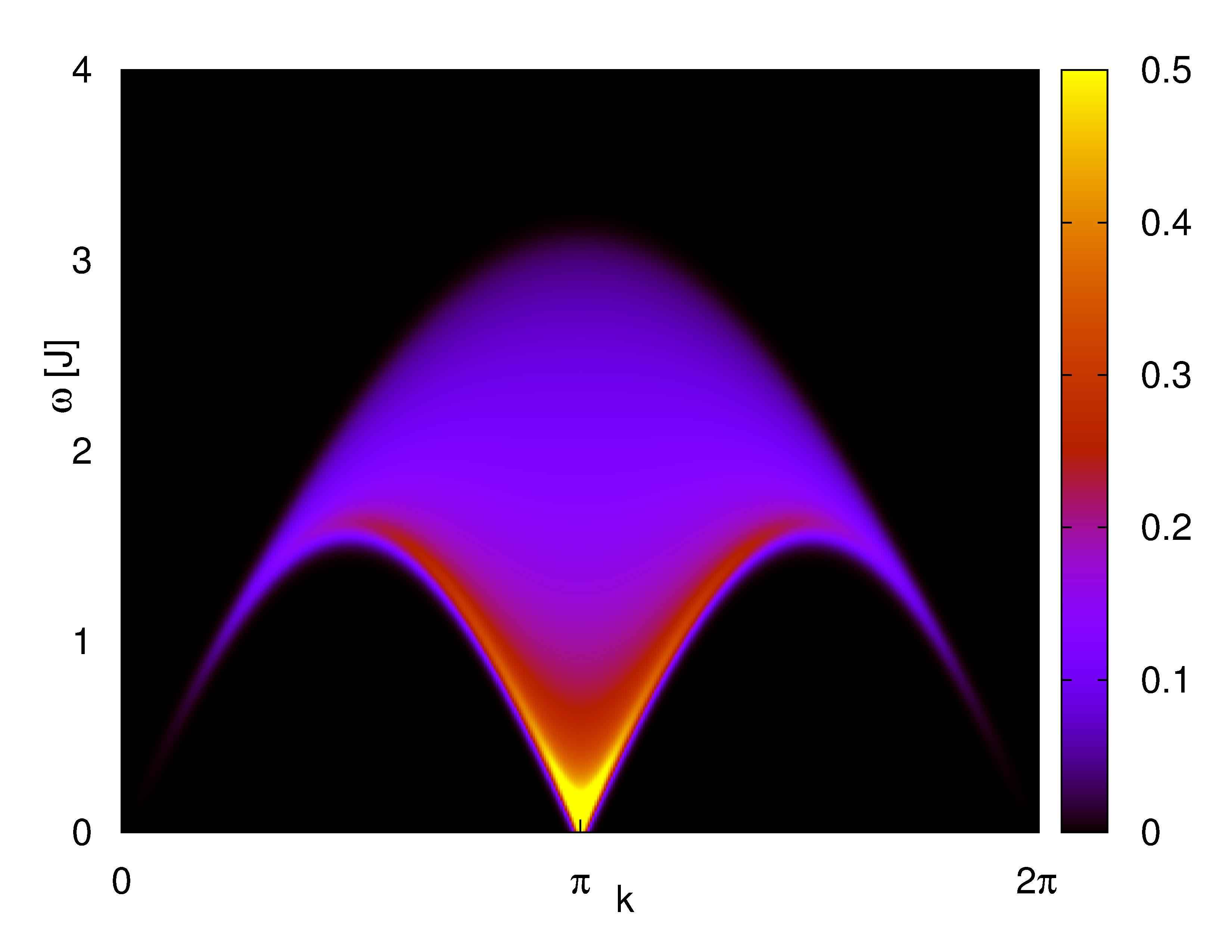}
\label{fig:2sp_DSF_SZZ_M200}
}
&\subfigure[$\;4$sp$2\infty$]{
\includegraphics[width=0.3\textwidth]{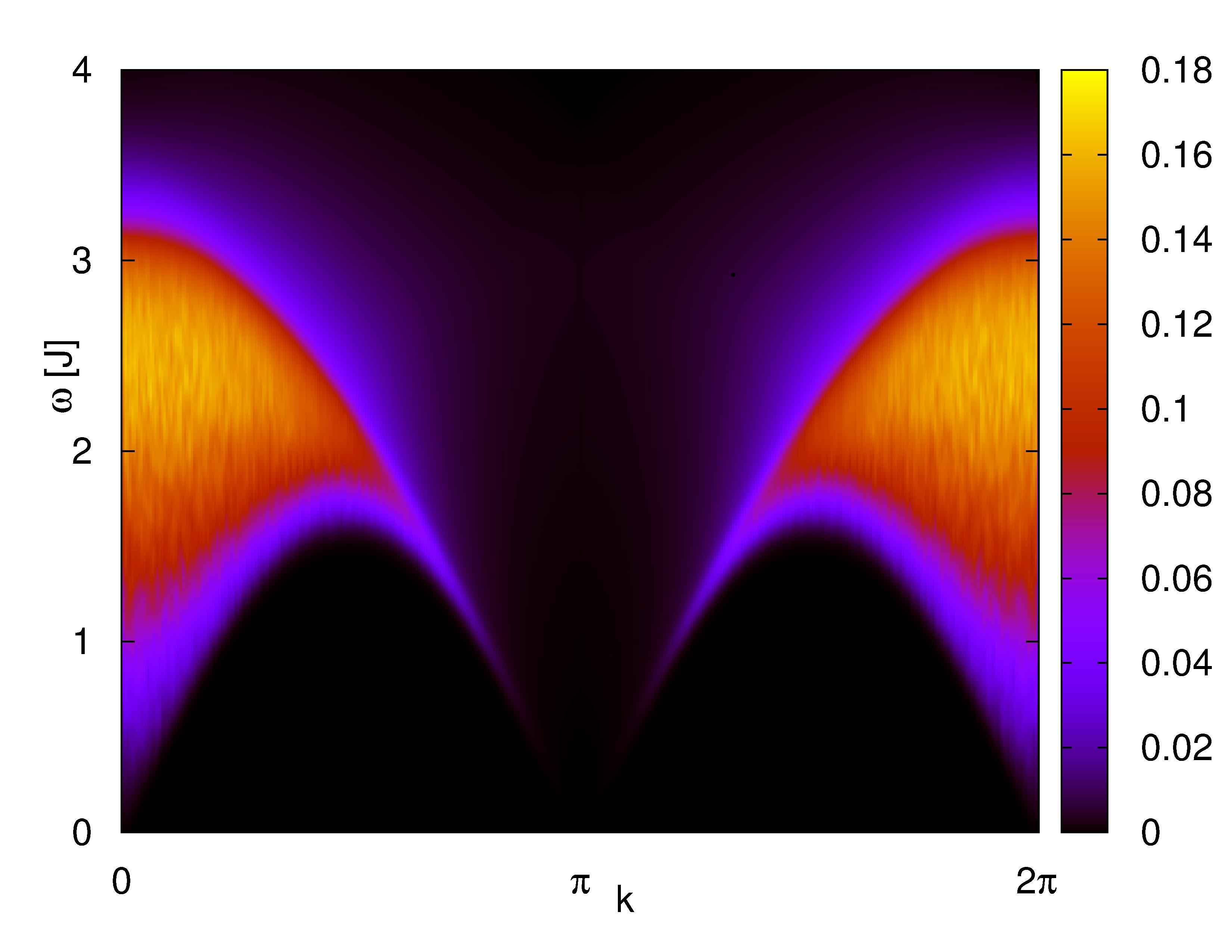}
\label{fig:4sp2inf_DSF_SZZ_M200}
}
\end{tabular}
\caption{Separated figures of the $S_c^{4z}(q,\omega)$ DSF with $N=400$ corresponding to types of excitation in the labels. The magnetization is $M=150$ in \subref{fig:1p_DSF_SZZ_M150}, \subref{fig:2p_DSF_SZZ_M150} and \subref{fig:1s_DSF_SZZ_M150} and $M=200$ for \subref{fig:2sp_DSF_SZZ_M200} and \subref{fig:4sp2inf_DSF_SZZ_M200}.}
\label{fig:Excit_type_DSF_SZZ}
\end{figure}

\end{widetext}
\section{Conclusion}
\label{sec:conclusion}

Throughout this article we have described the construction of the $S^-_jS^-_{j+1}$ and $S^z_jS^z_{j+1}$ form factors and we have used these results to formulate the adjacent spin operator correlation functions $S^{--++} (q, \omega)$ and $S^{4z}(q,\omega)$. The sum rules of each DSF have been given and  we have numerically evaluated the $S^{--++} (q, \omega)$ and $S^{4z}(q,\omega)$ DSF in an isotropic spin chain with $400$ sites in several magnetic fields. Among other results, we find a large contribution of the $4$-spinon states in the $S^{4z}(q,\omega)$ DSF. Moreover, we have shown that in some regions of the momentum-energy space, the signal can be attributed to $4$-spinon excitations and clearly distinguished from the $2$-spinon signal. There is presently, no experimental setup which allows one to measure such an observable but we can trust that, with the progresses in realizations of spin chains with Bose gases in optical lattices \cite{2011_Weitenberg_nature,2011_chen_preprint,2011_simon_nature}, measurements of adjacent spin operator correlation functions should be possible.

Besides, the size scaling of $S^-_jS^-_{j+1}$ and $S^z_jS^z_{j+1}$ FFs at the spectrum lower boundary could be compared to results of the low energy effective field theory \cite{2010_Shashi_preprint} for $\omega\ll 1$ and to the predictions of the non-linear Luttinger Liquid method \cite{2011_imambekov_preprint} for higher energies.

More generally, the results presented can be extended in different manners.
The FFs of other operators $S^a_iS^b_{i+1}$ with $a,b = z,+,-$ can be constructed. FFs including more next neighboring spin operators should in principle also be possible keeping in mind that every $S_j^z$ or $S_j^+$ operator adds a factor $N$ or $N^2$, respectively, to the total computation load. Moreover, the present formula for the FFs have been expressed for a general anisotropy and can easily be implemented for the evaluation of $S^{4z}(q,\omega)$ or  $S^{--++}(q,\omega)$ in the massless and massive regime, respectively $\Delta<1$ and $\Delta>1$. 

J. M. and J.-S. C. acknowledge support from the Foundation for Fundamental Research on Matter (FOM),
which is part of the Netherlands Organisation for Scientific Research (NWO).

\begin{appendix}
\end{appendix}

\bibliographystyle{apsrev}
\bibliography{references}

\begin{thebibliography}{57}
\expandafter\ifx\csname natexlab\endcsname\relax\def\natexlab#1{#1}\fi
\expandafter\ifx\csname bibnamefont\endcsname\relax
  \def\bibnamefont#1{#1}\fi
\expandafter\ifx\csname bibfnamefont\endcsname\relax
  \def\bibfnamefont#1{#1}\fi
\expandafter\ifx\csname citenamefont\endcsname\relax
  \def\citenamefont#1{#1}\fi
\expandafter\ifx\csname url\endcsname\relax
  \def\url#1{\texttt{#1}}\fi
\expandafter\ifx\csname urlprefix\endcsname\relax\def\urlprefix{URL }\fi
\providecommand{\bibinfo}[2]{#2}
\providecommand{\eprint}[2][]{\url{#2}}

\bibitem[{\citenamefont{Giamarchi}(2004)}]{giamarchibook}
\bibinfo{author}{\bibfnamefont{T.}~\bibnamefont{Giamarchi}},
  \emph{\bibinfo{title}{Quantum Physics in One Dimension}}
  (\bibinfo{publisher}{Oxford University Press}, \bibinfo{year}{2004}).

\bibitem[{\citenamefont{Sutherland}(2004)}]{sutherlandbook}
\bibinfo{author}{\bibfnamefont{B.}~\bibnamefont{Sutherland}},
  \emph{\bibinfo{title}{Beautiful models: 70 years of exactly solved quantum
  many-body problems}} (\bibinfo{publisher}{World Scientific},
  \bibinfo{year}{2004}).

\bibitem[{\citenamefont{Heisenberg}(1928)}]{1928_heisenberg_zp_49}
\bibinfo{author}{\bibfnamefont{W.}~\bibnamefont{Heisenberg}},
  \bibinfo{journal}{Z. Phys.} \textbf{\bibinfo{volume}{49}},
  \bibinfo{pages}{619} (\bibinfo{year}{1928}).

\bibitem[{\citenamefont{Nagler et~al.}(1991)\citenamefont{Nagler, Tennant,
  Cowley, Perring, and Satija}}]{1991_nagler_prb}
\bibinfo{author}{\bibfnamefont{S.~E.} \bibnamefont{Nagler}},
  \bibinfo{author}{\bibfnamefont{D.~A.} \bibnamefont{Tennant}},
  \bibinfo{author}{\bibfnamefont{R.~A.} \bibnamefont{Cowley}},
  \bibinfo{author}{\bibfnamefont{T.~G.} \bibnamefont{Perring}},
  \bibnamefont{and} \bibinfo{author}{\bibfnamefont{S.~K.}
  \bibnamefont{Satija}}, \bibinfo{journal}{Phys. Rev. B}
  \textbf{\bibinfo{volume}{44}}, \bibinfo{pages}{12361} (\bibinfo{year}{1991}).

\bibitem[{\citenamefont{Lake et~al.}(2005)\citenamefont{Lake, Tennant, Frost,
  and Nagler}}]{2005_lake_natmat}
\bibinfo{author}{\bibfnamefont{B.}~\bibnamefont{Lake}},
  \bibinfo{author}{\bibfnamefont{D.~A.} \bibnamefont{Tennant}},
  \bibinfo{author}{\bibfnamefont{C.~D.} \bibnamefont{Frost}}, \bibnamefont{and}
  \bibinfo{author}{\bibfnamefont{S.~E.} \bibnamefont{Nagler}},
  \bibinfo{journal}{Nature Materials} \textbf{\bibinfo{volume}{4}},
  \bibinfo{pages}{329} (\bibinfo{year}{2005}).

\bibitem[{\citenamefont{Stone et~al.}(2003)\citenamefont{Stone, Reich, Broholm,
  Lefmann, Rischel, Landee, and Turnbull}}]{2003_stone_prl}
\bibinfo{author}{\bibfnamefont{M.~B.} \bibnamefont{Stone}},
  \bibinfo{author}{\bibfnamefont{D.~H.} \bibnamefont{Reich}},
  \bibinfo{author}{\bibfnamefont{C.}~\bibnamefont{Broholm}},
  \bibinfo{author}{\bibfnamefont{K.}~\bibnamefont{Lefmann}},
  \bibinfo{author}{\bibfnamefont{C.}~\bibnamefont{Rischel}},
  \bibinfo{author}{\bibfnamefont{C.~P.} \bibnamefont{Landee}},
  \bibnamefont{and} \bibinfo{author}{\bibfnamefont{M.~M.}
  \bibnamefont{Turnbull}}, \bibinfo{journal}{Phys. Rev. Lett.}
  \textbf{\bibinfo{volume}{91}}, \bibinfo{pages}{037205}
  (\bibinfo{year}{2003}).

\bibitem[{\citenamefont{Zaliznyak et~al.}(2004)\citenamefont{Zaliznyak, Woo,
  Perring, Broholm, Frost, and Takagi}}]{2004_zaliznyak_prl}
\bibinfo{author}{\bibfnamefont{I.~A.} \bibnamefont{Zaliznyak}},
  \bibinfo{author}{\bibfnamefont{H.}~\bibnamefont{Woo}},
  \bibinfo{author}{\bibfnamefont{T.~G.} \bibnamefont{Perring}},
  \bibinfo{author}{\bibfnamefont{C.~L.} \bibnamefont{Broholm}},
  \bibinfo{author}{\bibfnamefont{C.~D.} \bibnamefont{Frost}}, \bibnamefont{and}
  \bibinfo{author}{\bibfnamefont{H.}~\bibnamefont{Takagi}},
  \bibinfo{journal}{Phys. Rev. Lett.} \textbf{\bibinfo{volume}{93}},
  \bibinfo{pages}{087202} (\bibinfo{year}{2004}).

\bibitem[{\citenamefont{Walters et~al.}(2009)\citenamefont{Walters, Perring,
  Caux, Savici, Gu, Lee, Ku, and Zaliznyak}}]{2009_walters_natphys_5}
\bibinfo{author}{\bibfnamefont{A.~C.} \bibnamefont{Walters}},
  \bibinfo{author}{\bibfnamefont{T.~G.} \bibnamefont{Perring}},
  \bibinfo{author}{\bibfnamefont{J.-S.} \bibnamefont{Caux}},
  \bibinfo{author}{\bibfnamefont{A.~T.} \bibnamefont{Savici}},
  \bibinfo{author}{\bibfnamefont{G.~D.} \bibnamefont{Gu}},
  \bibinfo{author}{\bibfnamefont{C.-C.} \bibnamefont{Lee}},
  \bibinfo{author}{\bibfnamefont{W.}~\bibnamefont{Ku}}, \bibnamefont{and}
  \bibinfo{author}{\bibfnamefont{I.~A.} \bibnamefont{Zaliznyak}},
  \bibinfo{journal}{Nat. Phys.} \textbf{\bibinfo{volume}{5}},
  \bibinfo{pages}{867} (\bibinfo{year}{2009}).

\bibitem[{\citenamefont{Nilsen et~al.}(2008)\citenamefont{Nilsen, R{\o}nnow,
  L{\"a}uchli, Fabbiani, Sanchez-Benitez, Kamenev, and
  Harrison}}]{2008_nilson_chemm}
\bibinfo{author}{\bibfnamefont{G.~J.} \bibnamefont{Nilsen}},
  \bibinfo{author}{\bibfnamefont{H.~M.} \bibnamefont{R{\o}nnow}},
  \bibinfo{author}{\bibfnamefont{A.~M.} \bibnamefont{L{\"a}uchli}},
  \bibinfo{author}{\bibfnamefont{F.~P.~A.} \bibnamefont{Fabbiani}},
  \bibinfo{author}{\bibfnamefont{J.}~\bibnamefont{Sanchez-Benitez}},
  \bibinfo{author}{\bibfnamefont{K.~V.} \bibnamefont{Kamenev}},
  \bibnamefont{and} \bibinfo{author}{\bibfnamefont{A.}~\bibnamefont{Harrison}},
  \bibinfo{journal}{Chem. Mat.} \textbf{\bibinfo{volume}{20}},
  \bibinfo{pages}{8} (\bibinfo{year}{2008}).

\bibitem[{\citenamefont{Enderle et~al.}(2005)\citenamefont{Enderle, Mukherjee,
  Fåk, Kremer, Broto, Rosner, Drechsler, Richter, Malek, Prokofiev
  et~al.}}]{2005_enderle_epl}
\bibinfo{author}{\bibfnamefont{M.}~\bibnamefont{Enderle}},
  \bibinfo{author}{\bibfnamefont{C.}~\bibnamefont{Mukherjee}},
  \bibinfo{author}{\bibfnamefont{B.}~\bibnamefont{Fåk}},
  \bibinfo{author}{\bibfnamefont{R.~K.} \bibnamefont{Kremer}},
  \bibinfo{author}{\bibfnamefont{J.-M.} \bibnamefont{Broto}},
  \bibinfo{author}{\bibfnamefont{H.}~\bibnamefont{Rosner}},
  \bibinfo{author}{\bibfnamefont{S.-L.} \bibnamefont{Drechsler}},
  \bibinfo{author}{\bibfnamefont{J.}~\bibnamefont{Richter}},
  \bibinfo{author}{\bibfnamefont{J.}~\bibnamefont{Malek}},
  \bibinfo{author}{\bibfnamefont{A.}~\bibnamefont{Prokofiev}},
  \bibnamefont{et~al.}, \bibinfo{journal}{Europhys. Lett.}
  \textbf{\bibinfo{volume}{70}}, \bibinfo{pages}{237} (\bibinfo{year}{2005}).

\bibitem[{\citenamefont{Faddeev and Takhtajan}(1981)}]{1981_faddeev_pla_85}
\bibinfo{author}{\bibfnamefont{L.~D.} \bibnamefont{Faddeev}} \bibnamefont{and}
  \bibinfo{author}{\bibfnamefont{L.~A.} \bibnamefont{Takhtajan}},
  \bibinfo{journal}{Phys. Lett. A} \textbf{\bibinfo{volume}{85}},
  \bibinfo{pages}{375} (\bibinfo{year}{1981}).

\bibitem[{\citenamefont{Bethe}(1931)}]{1931_bethe_zp_71}
\bibinfo{author}{\bibfnamefont{H.}~\bibnamefont{Bethe}}, \bibinfo{journal}{Z.
  Phys.} \textbf{\bibinfo{volume}{71}}, \bibinfo{pages}{205}
  (\bibinfo{year}{1931}).

\bibitem[{\citenamefont{Orbach}(1958)}]{1958_orbach_pr}
\bibinfo{author}{\bibfnamefont{R.}~\bibnamefont{Orbach}},
  \bibinfo{journal}{Phys. Rev.} \textbf{\bibinfo{volume}{112}},
  \bibinfo{pages}{309} (\bibinfo{year}{1958}).

\bibitem[{\citenamefont{Yang and Yang}(1966{\natexlab{a}})}]{1966_yang_pr1}
\bibinfo{author}{\bibfnamefont{C.~N.} \bibnamefont{Yang}} \bibnamefont{and}
  \bibinfo{author}{\bibfnamefont{C.~P.} \bibnamefont{Yang}},
  \bibinfo{journal}{Phys. Rev.} \textbf{\bibinfo{volume}{150}},
  \bibinfo{pages}{321} (\bibinfo{year}{1966}{\natexlab{a}}).

\bibitem[{\citenamefont{Yang and Yang}(1966{\natexlab{b}})}]{1966_yang_pr2}
\bibinfo{author}{\bibfnamefont{C.~N.} \bibnamefont{Yang}} \bibnamefont{and}
  \bibinfo{author}{\bibfnamefont{C.~P.} \bibnamefont{Yang}},
  \bibinfo{journal}{Phys. Rev.} \textbf{\bibinfo{volume}{150}},
  \bibinfo{pages}{327} (\bibinfo{year}{1966}{\natexlab{b}}).

\bibitem[{\citenamefont{Yang and Yang}(1966{\natexlab{c}})}]{1966_yang_pr3}
\bibinfo{author}{\bibfnamefont{C.~N.} \bibnamefont{Yang}} \bibnamefont{and}
  \bibinfo{author}{\bibfnamefont{C.~P.} \bibnamefont{Yang}},
  \bibinfo{journal}{Phys. Rev.} \textbf{\bibinfo{volume}{151}},
  \bibinfo{pages}{258} (\bibinfo{year}{1966}{\natexlab{c}}).

\bibitem[{\citenamefont{Kurmann et~al.}(1982)\citenamefont{Kurmann, Thomas, and
  Mueller}}]{1982_Kurmann_Phy}
\bibinfo{author}{\bibfnamefont{J.}~\bibnamefont{Kurmann}},
  \bibinfo{author}{\bibfnamefont{H.}~\bibnamefont{Thomas}}, \bibnamefont{and}
  \bibinfo{author}{\bibfnamefont{G.}~\bibnamefont{Mueller}},
  \bibinfo{journal}{Physica A} \textbf{\bibinfo{volume}{112}},
  \bibinfo{pages}{235 } (\bibinfo{year}{1982}).

\bibitem[{\citenamefont{Jimbo and Miwa}(1995)}]{jimbobook}
\bibinfo{author}{\bibfnamefont{M.}~\bibnamefont{Jimbo}} \bibnamefont{and}
  \bibinfo{author}{\bibfnamefont{T.}~\bibnamefont{Miwa}},
  \emph{\bibinfo{title}{Algebraic Analysis of Solvable Lattice Models}}
  (\bibinfo{publisher}{Providence, RI: AMS}, \bibinfo{year}{1995}).

\bibitem[{\citenamefont{Korepin et~al.}(1993)\citenamefont{Korepin, Bogoliubov,
  and Izergin}}]{korepinbook}
\bibinfo{author}{\bibfnamefont{V.~E.} \bibnamefont{Korepin}},
  \bibinfo{author}{\bibfnamefont{N.~M.} \bibnamefont{Bogoliubov}},
  \bibnamefont{and} \bibinfo{author}{\bibfnamefont{A.~G.}
  \bibnamefont{Izergin}}, \emph{\bibinfo{title}{Quantum Inverse Scattering
  Method and Correlation Functions}} (\bibinfo{publisher}{Cambridge Univ.
  Press}, \bibinfo{year}{1993}).

\bibitem[{\citenamefont{Karbach et~al.}(1997)\citenamefont{Karbach, M\"uller,
  Bougourzi, Fledderjohann, and M\"utter}}]{1997_karbach_prb_55}
\bibinfo{author}{\bibfnamefont{M.}~\bibnamefont{Karbach}},
  \bibinfo{author}{\bibfnamefont{G.}~\bibnamefont{M\"uller}},
  \bibinfo{author}{\bibfnamefont{A.~H.} \bibnamefont{Bougourzi}},
  \bibinfo{author}{\bibfnamefont{A.}~\bibnamefont{Fledderjohann}},
  \bibnamefont{and} \bibinfo{author}{\bibfnamefont{K.-H.}
  \bibnamefont{M\"utter}}, \bibinfo{journal}{Phys. Rev. B}
  \textbf{\bibinfo{volume}{55}}, \bibinfo{pages}{12510} (\bibinfo{year}{1997}).

\bibitem[{\citenamefont{Kitanine et~al.}(1999)\citenamefont{Kitanine, Maillet,
  and Terras}}]{1999_kitanine_npb_554}
\bibinfo{author}{\bibfnamefont{N.}~\bibnamefont{Kitanine}},
  \bibinfo{author}{\bibfnamefont{J.~M.} \bibnamefont{Maillet}},
  \bibnamefont{and} \bibinfo{author}{\bibfnamefont{V.}~\bibnamefont{Terras}},
  \bibinfo{journal}{Nucl. Phys. B} \textbf{\bibinfo{volume}{554}},
  \bibinfo{pages}{647 } (\bibinfo{year}{1999}).

\bibitem[{\citenamefont{Kitanine et~al.}(2000)\citenamefont{Kitanine, Maillet,
  and Terras}}]{2000_kitanine_npb_567}
\bibinfo{author}{\bibfnamefont{N.}~\bibnamefont{Kitanine}},
  \bibinfo{author}{\bibfnamefont{J.~M.} \bibnamefont{Maillet}},
  \bibnamefont{and} \bibinfo{author}{\bibfnamefont{V.}~\bibnamefont{Terras}},
  \bibinfo{journal}{Nucl. Phys. B} \textbf{\bibinfo{volume}{567}},
  \bibinfo{pages}{554 } (\bibinfo{year}{2000}).

\bibitem[{\citenamefont{Caux et~al.}(2003)\citenamefont{Caux, Essler, and
  L\"ow}}]{2003_caux_prb}
\bibinfo{author}{\bibfnamefont{J.-S.} \bibnamefont{Caux}},
  \bibinfo{author}{\bibfnamefont{F.~H.~L.} \bibnamefont{Essler}},
  \bibnamefont{and} \bibinfo{author}{\bibfnamefont{U.}~\bibnamefont{L\"ow}},
  \bibinfo{journal}{Phys. Rev. B} \textbf{\bibinfo{volume}{68}},
  \bibinfo{pages}{134431} (\bibinfo{year}{2003}).

\bibitem[{\citenamefont{Caux and Maillet}(2005)}]{2005_caux_prl_95}
\bibinfo{author}{\bibfnamefont{J.-S.} \bibnamefont{Caux}} \bibnamefont{and}
  \bibinfo{author}{\bibfnamefont{J.~M.} \bibnamefont{Maillet}},
  \bibinfo{journal}{Phys. Rev. Lett.} \textbf{\bibinfo{volume}{95}},
  \bibinfo{pages}{077201} (\bibinfo{year}{2005}).

\bibitem[{\citenamefont{Caux et~al.}(2005)\citenamefont{Caux, Hagemans, and
  Maillet}}]{2005_caux_jstat_p09003}
\bibinfo{author}{\bibfnamefont{J.-S.} \bibnamefont{Caux}},
  \bibinfo{author}{\bibfnamefont{R.}~\bibnamefont{Hagemans}}, \bibnamefont{and}
  \bibinfo{author}{\bibfnamefont{J.~M.} \bibnamefont{Maillet}},
  \bibinfo{journal}{J. Stat. Mech.} p. \bibinfo{pages}{P09003}
  (\bibinfo{year}{2005}).

\bibitem[{\citenamefont{Pereira et~al.}(2007)\citenamefont{Pereira, Sirker,
  Caux, Hagemans, Maillet, White, and Affleck}}]{2007_pereira_jstat}
\bibinfo{author}{\bibfnamefont{R.~G.} \bibnamefont{Pereira}},
  \bibinfo{author}{\bibfnamefont{J.}~\bibnamefont{Sirker}},
  \bibinfo{author}{\bibfnamefont{J.-S.} \bibnamefont{Caux}},
  \bibinfo{author}{\bibfnamefont{R.}~\bibnamefont{Hagemans}},
  \bibinfo{author}{\bibfnamefont{J.~M.} \bibnamefont{Maillet}},
  \bibinfo{author}{\bibfnamefont{S.~R.} \bibnamefont{White}}, \bibnamefont{and}
  \bibinfo{author}{\bibfnamefont{I.}~\bibnamefont{Affleck}},
  \bibinfo{journal}{J. Stat. Mech.} p. \bibinfo{pages}{P08022}
  (\bibinfo{year}{2007}).

\bibitem[{\citenamefont{Caux et~al.}(2008)\citenamefont{Caux, Mossel, and
  {P{\'e}rez Castillo}}}]{2008_caux_jstat_p08006}
\bibinfo{author}{\bibfnamefont{J.-S.} \bibnamefont{Caux}},
  \bibinfo{author}{\bibfnamefont{J.}~\bibnamefont{Mossel}}, \bibnamefont{and}
  \bibinfo{author}{\bibfnamefont{I.}~\bibnamefont{{P{\'e}rez Castillo}}},
  \bibinfo{journal}{J. Stat. Mech.} p. \bibinfo{pages}{P08006}
  (\bibinfo{year}{2008}).

\bibitem[{\citenamefont{Caux et~al.}(2011)\citenamefont{Caux, Konno, Sorrell,
  and Weston}}]{2011_caux_prl}
\bibinfo{author}{\bibfnamefont{J.-S.} \bibnamefont{Caux}},
  \bibinfo{author}{\bibfnamefont{H.}~\bibnamefont{Konno}},
  \bibinfo{author}{\bibfnamefont{M.}~\bibnamefont{Sorrell}}, \bibnamefont{and}
  \bibinfo{author}{\bibfnamefont{R.}~\bibnamefont{Weston}},
  \bibinfo{journal}{Phys. Rev. Lett.} \textbf{\bibinfo{volume}{106}},
  \bibinfo{pages}{217203} (\bibinfo{year}{2011}).

\bibitem[{\citenamefont{Klauser et~al.}(2011)\citenamefont{Klauser, Mossel,
  Caux, and van~den Brink}}]{2011_klauser_prl}
\bibinfo{author}{\bibfnamefont{A.}~\bibnamefont{Klauser}},
  \bibinfo{author}{\bibfnamefont{J.}~\bibnamefont{Mossel}},
  \bibinfo{author}{\bibfnamefont{J.-S.} \bibnamefont{Caux}}, \bibnamefont{and}
  \bibinfo{author}{\bibfnamefont{J.}~\bibnamefont{van~den Brink}},
  \bibinfo{journal}{Phys. Rev. Lett.} \textbf{\bibinfo{volume}{106}},
  \bibinfo{pages}{157205} (\bibinfo{year}{2011}).

\bibitem[{\citenamefont{{Zotos} and {Prelovsek}}(2003)}]{2003_zotos_arx}
\bibinfo{author}{\bibfnamefont{X.}~\bibnamefont{{Zotos}}} \bibnamefont{and}
  \bibinfo{author}{\bibfnamefont{P.}~\bibnamefont{{Prelovsek}}},
  \bibinfo{journal}{ArXiv}  (\bibinfo{year}{2003}), \eprint{cond-mat/0304630}.

\bibitem[{\citenamefont{Narozhny et~al.}(1998)\citenamefont{Narozhny, Millis,
  and Andrei}}]{1998_narozhny_prb}
\bibinfo{author}{\bibfnamefont{B.~N.} \bibnamefont{Narozhny}},
  \bibinfo{author}{\bibfnamefont{A.~J.} \bibnamefont{Millis}},
  \bibnamefont{and} \bibinfo{author}{\bibfnamefont{N.}~\bibnamefont{Andrei}},
  \bibinfo{journal}{Phys. Rev. B} \textbf{\bibinfo{volume}{58}},
  \bibinfo{pages}{R2921} (\bibinfo{year}{1998}).

\bibitem[{\citenamefont{Sirker et~al.}(2011)\citenamefont{Sirker, Pereira, and
  Affleck}}]{2011_sirker_prb}
\bibinfo{author}{\bibfnamefont{J.}~\bibnamefont{Sirker}},
  \bibinfo{author}{\bibfnamefont{R.~G.} \bibnamefont{Pereira}},
  \bibnamefont{and} \bibinfo{author}{\bibfnamefont{I.}~\bibnamefont{Affleck}},
  \bibinfo{journal}{Phys. Rev. B} \textbf{\bibinfo{volume}{83}},
  \bibinfo{pages}{035115} (\bibinfo{year}{2011}).

\bibitem[{\citenamefont{Steinigeweg et~al.}(2011)\citenamefont{Steinigeweg,
  Langer, Heidrich-Meisner, McCulloch, and Brenig}}]{2011_steinigeweg_prl}
\bibinfo{author}{\bibfnamefont{R.}~\bibnamefont{Steinigeweg}},
  \bibinfo{author}{\bibfnamefont{S.}~\bibnamefont{Langer}},
  \bibinfo{author}{\bibfnamefont{F.}~\bibnamefont{Heidrich-Meisner}},
  \bibinfo{author}{\bibfnamefont{I.~P.} \bibnamefont{McCulloch}},
  \bibnamefont{and} \bibinfo{author}{\bibfnamefont{W.}~\bibnamefont{Brenig}},
  \bibinfo{journal}{Phys. Rev. Lett.} \textbf{\bibinfo{volume}{106}},
  \bibinfo{pages}{160602} (\bibinfo{year}{2011}).

\bibitem[{\citenamefont{van Hoogdalem and Loss}(2011)}]{2011_hoogdalem_prb}
\bibinfo{author}{\bibfnamefont{K.~A.} \bibnamefont{van Hoogdalem}}
  \bibnamefont{and} \bibinfo{author}{\bibfnamefont{D.}~\bibnamefont{Loss}},
  \bibinfo{journal}{Phys. Rev. B} \textbf{\bibinfo{volume}{84}},
  \bibinfo{pages}{024402} (\bibinfo{year}{2011}).

\bibitem[{\citenamefont{Zanardi and Paunkovi\ifmmode~\acute{c}\else
  \'{c}\fi{}}(2006)}]{2006_zanardi_pre}
\bibinfo{author}{\bibfnamefont{P.}~\bibnamefont{Zanardi}} \bibnamefont{and}
  \bibinfo{author}{\bibfnamefont{N.}~\bibnamefont{Paunkovi\ifmmode~\acute{c}\e%
lse \'{c}\fi{}}}, \bibinfo{journal}{Phys. Rev. E}
  \textbf{\bibinfo{volume}{74}}, \bibinfo{pages}{031123}
  (\bibinfo{year}{2006}).

\bibitem[{\citenamefont{Albuquerque et~al.}(2010)\citenamefont{Albuquerque,
  Alet, Sire, and Capponi}}]{2010_albuquerque_prb}
\bibinfo{author}{\bibfnamefont{A.~F.} \bibnamefont{Albuquerque}},
  \bibinfo{author}{\bibfnamefont{F.}~\bibnamefont{Alet}},
  \bibinfo{author}{\bibfnamefont{C.}~\bibnamefont{Sire}}, \bibnamefont{and}
  \bibinfo{author}{\bibfnamefont{S.}~\bibnamefont{Capponi}},
  \bibinfo{journal}{Phys. Rev. B} \textbf{\bibinfo{volume}{81}},
  \bibinfo{pages}{064418} (\bibinfo{year}{2010}).

\bibitem[{\citenamefont{Giamarchi and Schulz}(1989)}]{1989_giamarchi_prb_39}
\bibinfo{author}{\bibfnamefont{T.}~\bibnamefont{Giamarchi}} \bibnamefont{and}
  \bibinfo{author}{\bibfnamefont{H.~J.} \bibnamefont{Schulz}},
  \bibinfo{journal}{Phys. Rev. B} \textbf{\bibinfo{volume}{39}},
  \bibinfo{pages}{4620} (\bibinfo{year}{1989}).

\bibitem[{\citenamefont{Hikihara and Furusaki}(2004)}]{2004_hikihara_prb}
\bibinfo{author}{\bibfnamefont{T.}~\bibnamefont{Hikihara}} \bibnamefont{and}
  \bibinfo{author}{\bibfnamefont{A.}~\bibnamefont{Furusaki}},
  \bibinfo{journal}{Phys. Rev. B} \textbf{\bibinfo{volume}{69}},
  \bibinfo{pages}{064427} (\bibinfo{year}{2004}).

\bibitem[{\citenamefont{Takahashi}(1971)}]{1971_takahashi_ptp}
\bibinfo{author}{\bibfnamefont{M.}~\bibnamefont{Takahashi}},
  \bibinfo{journal}{Prog. Theor. Phys.} \textbf{\bibinfo{volume}{46}},
  \bibinfo{pages}{401} (\bibinfo{year}{1971}).

\bibitem[{\citenamefont{Fujita et~al.}(2003)\citenamefont{Fujita, Kobayashi,
  and Takahashi}}]{2003_fujita_jpa}
\bibinfo{author}{\bibfnamefont{T.}~\bibnamefont{Fujita}},
  \bibinfo{author}{\bibfnamefont{T.}~\bibnamefont{Kobayashi}},
  \bibnamefont{and}
  \bibinfo{author}{\bibfnamefont{H.}~\bibnamefont{Takahashi}},
  \bibinfo{journal}{J. Phys. A} \textbf{\bibinfo{volume}{36}},
  \bibinfo{pages}{1553} (\bibinfo{year}{2003}).

\bibitem[{\citenamefont{Hagemans and Caux}(2007)}]{2007_hagemans_jpa}
\bibinfo{author}{\bibfnamefont{R.}~\bibnamefont{Hagemans}} \bibnamefont{and}
  \bibinfo{author}{\bibfnamefont{J.-S.} \bibnamefont{Caux}},
  \bibinfo{journal}{J. Phys. A} \textbf{\bibinfo{volume}{40}},
  \bibinfo{pages}{14605} (\bibinfo{year}{2007}).

\bibitem[{\citenamefont{Gaudin}(1983)}]{gaudinbook}
\bibinfo{author}{\bibfnamefont{M.}~\bibnamefont{Gaudin}},
  \emph{\bibinfo{title}{La fonction d'onde de {B}ethe}}
  (\bibinfo{publisher}{Masson, Paris}, \bibinfo{year}{1983}).

\bibitem[{\citenamefont{Gaudin et~al.}(1981)\citenamefont{Gaudin, McCoy, and
  Wu}}]{1981_gaudin_prd_23}
\bibinfo{author}{\bibfnamefont{M.}~\bibnamefont{Gaudin}},
  \bibinfo{author}{\bibfnamefont{B.~M.} \bibnamefont{McCoy}}, \bibnamefont{and}
  \bibinfo{author}{\bibfnamefont{T.~T.} \bibnamefont{Wu}},
  \bibinfo{journal}{Phys. Rev. D} \textbf{\bibinfo{volume}{23}},
  \bibinfo{pages}{417} (\bibinfo{year}{1981}).

\bibitem[{\citenamefont{Slavnov}(1989)}]{1989_slavnov_tmp_79}
\bibinfo{author}{\bibfnamefont{N.~A.} \bibnamefont{Slavnov}},
  \bibinfo{journal}{Teor. Mat. Fiz.} \textbf{\bibinfo{volume}{79}},
  \bibinfo{pages}{232} (\bibinfo{year}{1989}).

\bibitem[{\citenamefont{Korepin}(1982)}]{1982_korepin_cmp_86}
\bibinfo{author}{\bibfnamefont{V.~E.} \bibnamefont{Korepin}},
  \bibinfo{journal}{Comm. Math. Phys.} \textbf{\bibinfo{volume}{86}},
  \bibinfo{pages}{391} (\bibinfo{year}{1982}).

\bibitem[{\citenamefont{Kirillov and Korepin}(1988)}]{kirillov_jms_1988}
\bibinfo{author}{\bibfnamefont{A.~N.} \bibnamefont{Kirillov}} \bibnamefont{and}
  \bibinfo{author}{\bibfnamefont{V.~E.} \bibnamefont{Korepin}},
  \bibinfo{journal}{J. Math. Sc.} \textbf{\bibinfo{volume}{40}},
  \bibinfo{pages}{13} (\bibinfo{year}{1988}).

\bibitem[{\citenamefont{Sakurai}(1994)}]{sakurai_book}
\bibinfo{author}{\bibfnamefont{J.~J.} \bibnamefont{Sakurai}},
  \emph{\bibinfo{title}{Modern Quantum Mechanics}}
  (\bibinfo{publisher}{Addison-Wesley Publishing Company},
  \bibinfo{year}{1994}), \bibinfo{edition}{{R}evised} ed.

\bibitem[{\citenamefont{Sakai et~al.}(2003)\citenamefont{Sakai, Shiroishi,
  Nishiyama, and Takahashi}}]{2003_sakai_pre_67}
\bibinfo{author}{\bibfnamefont{K.}~\bibnamefont{Sakai}},
  \bibinfo{author}{\bibfnamefont{M.}~\bibnamefont{Shiroishi}},
  \bibinfo{author}{\bibfnamefont{Y.}~\bibnamefont{Nishiyama}},
  \bibnamefont{and}
  \bibinfo{author}{\bibfnamefont{M.}~\bibnamefont{Takahashi}},
  \bibinfo{journal}{Phys. Rev. E} \textbf{\bibinfo{volume}{67}},
  \bibinfo{pages}{065101} (\bibinfo{year}{2003}).

\bibitem[{\citenamefont{Caux}(2009)}]{2009_caux_jmp_50}
\bibinfo{author}{\bibfnamefont{J.-S.} \bibnamefont{Caux}}, \bibinfo{journal}{J.
  Math. Phys.} \textbf{\bibinfo{volume}{50}}, \bibinfo{eid}{095214}
  (\bibinfo{year}{2009}).

\bibitem[{\citenamefont{Kohno}(2009)}]{2009_kohno_prl_102}
\bibinfo{author}{\bibfnamefont{M.}~\bibnamefont{Kohno}},
  \bibinfo{journal}{Phys. Rev. Lett.} \textbf{\bibinfo{volume}{102}},
  \bibinfo{pages}{037203} (\bibinfo{year}{2009}).

\bibitem[{\citenamefont{{Ganahl} et~al.}(2011)\citenamefont{{Ganahl}, {Rabel},
  {Essler}, and {Evertz}}}]{2011_ganahl_arxiv}
\bibinfo{author}{\bibfnamefont{M.}~\bibnamefont{{Ganahl}}},
  \bibinfo{author}{\bibfnamefont{E.}~\bibnamefont{{Rabel}}},
  \bibinfo{author}{\bibfnamefont{F.~H.~L.} \bibnamefont{{Essler}}},
  \bibnamefont{and} \bibinfo{author}{\bibfnamefont{H.~G.}
  \bibnamefont{{Evertz}}}, \bibinfo{journal}{ArXiv}  (\bibinfo{year}{2011}),
  \eprint{cond-mat/1112.3355}.

\bibitem[{\citenamefont{Caux and Hagemans}(2006)}]{2006_caux_jstat_p12013}
\bibinfo{author}{\bibfnamefont{J.-S.} \bibnamefont{Caux}} \bibnamefont{and}
  \bibinfo{author}{\bibfnamefont{R.}~\bibnamefont{Hagemans}},
  \bibinfo{journal}{J. Stat. Mech.} p. \bibinfo{pages}{P12013}
  (\bibinfo{year}{2006}).

\bibitem[{\citenamefont{Weitenberg et~al.}(2011)\citenamefont{Weitenberg,
  Endres, Sherson, Cheneau, Schausz, Fukuhara, Bloch, and
  Kuhr}}]{2011_Weitenberg_nature}
\bibinfo{author}{\bibfnamefont{C.}~\bibnamefont{Weitenberg}},
  \bibinfo{author}{\bibfnamefont{M.}~\bibnamefont{Endres}},
  \bibinfo{author}{\bibfnamefont{J.~F.} \bibnamefont{Sherson}},
  \bibinfo{author}{\bibfnamefont{M.}~\bibnamefont{Cheneau}},
  \bibinfo{author}{\bibfnamefont{P.}~\bibnamefont{Schausz}},
  \bibinfo{author}{\bibfnamefont{T.}~\bibnamefont{Fukuhara}},
  \bibinfo{author}{\bibfnamefont{I.}~\bibnamefont{Bloch}}, \bibnamefont{and}
  \bibinfo{author}{\bibfnamefont{S.}~\bibnamefont{Kuhr}},
  \bibinfo{journal}{Nature} \textbf{\bibinfo{volume}{471}},
  \bibinfo{pages}{319} (\bibinfo{year}{2011}).

\bibitem[{\citenamefont{{Chen} et~al.}(2011)\citenamefont{{Chen},
  {Nascimb{\`e}ne}, {Aidelsburger}, {Atala}, {Trotzky}, and
  {Bloch}}}]{2011_chen_preprint}
\bibinfo{author}{\bibfnamefont{Y.-A.} \bibnamefont{{Chen}}},
  \bibinfo{author}{\bibfnamefont{S.}~\bibnamefont{{Nascimb{\`e}ne}}},
  \bibinfo{author}{\bibfnamefont{M.}~\bibnamefont{{Aidelsburger}}},
  \bibinfo{author}{\bibfnamefont{M.}~\bibnamefont{{Atala}}},
  \bibinfo{author}{\bibfnamefont{S.}~\bibnamefont{{Trotzky}}},
  \bibnamefont{and} \bibinfo{author}{\bibfnamefont{I.}~\bibnamefont{{Bloch}}},
  \bibinfo{journal}{ArXiv}  (\bibinfo{year}{2011}),
  \eprint{cond-mat/1104.1833}.

\bibitem[{\citenamefont{Simon et~al.}(2011)\citenamefont{Simon, Bakr, Ma, Tai,
  Preiss, and Greiner}}]{2011_simon_nature}
\bibinfo{author}{\bibfnamefont{J.}~\bibnamefont{Simon}},
  \bibinfo{author}{\bibfnamefont{W.~S.} \bibnamefont{Bakr}},
  \bibinfo{author}{\bibfnamefont{R.}~\bibnamefont{Ma}},
  \bibinfo{author}{\bibfnamefont{M.~E.} \bibnamefont{Tai}},
  \bibinfo{author}{\bibfnamefont{P.~M.} \bibnamefont{Preiss}},
  \bibnamefont{and} \bibinfo{author}{\bibfnamefont{M.}~\bibnamefont{Greiner}},
  \bibinfo{journal}{Nature} \textbf{\bibinfo{volume}{472}},
  \bibinfo{pages}{307} (\bibinfo{year}{2011}).

\bibitem[{\citenamefont{{Shashi} et~al.}(2010)\citenamefont{{Shashi}, {Panfil},
  {Caux}, and {Imambekov}}}]{2010_Shashi_preprint}
\bibinfo{author}{\bibfnamefont{A.}~\bibnamefont{{Shashi}}},
  \bibinfo{author}{\bibfnamefont{M.}~\bibnamefont{{Panfil}}},
  \bibinfo{author}{\bibfnamefont{J.-S.} \bibnamefont{{Caux}}},
  \bibnamefont{and}
  \bibinfo{author}{\bibfnamefont{A.}~\bibnamefont{{Imambekov}}},
  \bibinfo{journal}{ArXiv}  (\bibinfo{year}{2010}),
  \eprint{cond-mat/1010.2268}.

\bibitem[{\citenamefont{{Imambekov} et~al.}(2011)\citenamefont{{Imambekov},
  {Schmidt}, and {Glazman}}}]{2011_imambekov_preprint}
\bibinfo{author}{\bibfnamefont{A.}~\bibnamefont{{Imambekov}}},
  \bibinfo{author}{\bibfnamefont{T.~L.} \bibnamefont{{Schmidt}}},
  \bibnamefont{and} \bibinfo{author}{\bibfnamefont{L.~I.}
  \bibnamefont{{Glazman}}}, \bibinfo{journal}{ArXiv}  (\bibinfo{year}{2011}),
  \eprint{cond-mat/1110.1374}.

\end{thebibliography}

\end{document}